\documentclass[%
    preprint,
    superscriptaddress,
    nofootinbib,
    amsmath,amssymb,
    aps,
    prb,
    floatfix,
]{revtex4-2}

\usepackage{graphicx}               
\usepackage[caption=false]{subfig}
\usepackage{tabularx}
\usepackage{dcolumn}                
\usepackage{bm}                     
\usepackage{braket}
\usepackage{siunitx}

\usepackage[colorlinks,
            linkcolor=blue,
            citecolor=blue,
            filecolor=blue,
            urlcolor=blue,
            anchorcolor=blue
            ]{hyperref}               
            
\newcommand{\bk}{{\bm k}}
\newcommand{\br}{{\bm r}}
\newcommand{\bR}{{\bm R}}
\newcommand{\meV}{\si{\milli\electronvolt}}

\begin{document}

\title{Bloch's theorem in orbital-density-dependent functionals:\break Band structures from Koopmans spectral functionals}

\author{Riccardo De Gennaro}
\email{riccardo.degennaro@epfl.ch}
\affiliation{Theory and Simulations of Materials (THEOS), and National Centre for Computational Design and Discovery of Novel Materials (MARVEL), \'{E}cole Polytechnique F\'{e}d\'{e}rale de Lausanne, 1015 Lausanne, Switzerland}
\author{Nicola Colonna}
\affiliation{Theory and Simulations of Materials (THEOS), and National Centre for Computational Design and Discovery of Novel Materials (MARVEL), \'{E}cole Polytechnique F\'{e}d\'{e}rale de Lausanne, 1015 Lausanne, Switzerland}
\affiliation{Laboratory for Neutron Scattering and Imaging, Paul Scherrer Institut, 5232 Villigen PSI, Switzerland}
\author{Edward Linscott}
\affiliation{Theory and Simulations of Materials (THEOS), and National Centre for Computational Design and Discovery of Novel Materials (MARVEL), \'{E}cole Polytechnique F\'{e}d\'{e}rale de Lausanne, 1015 Lausanne, Switzerland}
\author{Nicola Marzari}
\affiliation{Theory and Simulations of Materials (THEOS), and National Centre for Computational Design and Discovery of Novel Materials (MARVEL), \'{E}cole Polytechnique F\'{e}d\'{e}rale de Lausanne, 1015 Lausanne, Switzerland}
\affiliation{Laboratory for Materials Simulations, Paul Scherrer Institut, 5232 Villigen PSI, Switzerland}

\date{\today}

\begin{abstract}

Koopmans-compliant functionals provide a novel orbital-density-dependent framework for an accurate evaluation of spectral properties; they are obtained by imposing a generalized piecewise-linearity condition on the total energy of the system with respect to the occupation of any orbital. In crystalline materials, due to the orbital-density-dependent nature of the functionals, minimization of the total energy to a ground state provides a set of minimizing variational orbitals that are localized, and thus break the periodicity of the underlying lattice. Despite this, we show that Bloch symmetry can be preserved and it is possible to describe the electronic states with a band-structure picture, thanks to the Wannier-like character of the variational orbitals. We also present a method to unfold and interpolate the electronic bands from supercell ($\Gamma$-point) calculations, which enables us to calculate full band structures with Koopmans-compliant functionals. The results obtained for a set of benchmark semiconductors and insulators show a very good agreement with state-of-the-art many-body perturbation theory and experiments, underscoring the reliability of these spectral functionals in predicting band structures.

\end{abstract}

\maketitle

\section{Introduction}

The advent of density-functional theory (DFT) has made it possible to calculate the fundamental properties of materials from first principles. Thanks to Hohenberg-Kohn theorems \cite{hohenberg_inhomogeneous_1964}, a system's Hamiltonian is fully determined by the electronic ground-state density and thus so too are all excited-state properties. However, a major challenge is to find ways to extract the desired features from a functional of the density. The auxiliary non-interacting Kohn-Sham (KS) system \cite{kohn_self-consistent_1965} provides a framework to approximate in reliable and straightforward ways ground-state densities and energies, but it does not provide an explicit description of excited-state properties (e.g. see Ref.~\cite{marzari_electronic-structure_2021} and references therein), the one exception being the highest-occupied (HO) state which corresponds in exact KS-DFT to the opposite of the ionization potential (IP) of the system \cite{perdew_density-functional_1982, almbladh_exact_1985}. On the other hand, the connection between the band structure of a crystal (i.e.\ its $\bk$-resolved photoemission spectrum of charged excitations) and the KS orbital energies is less straightforward. The KS potential is the variationally best local and static approximation to the self-energy \cite{casida_generalization_1995}, and it was recently pointed out that local and dynamical potentials can improve the description of spectral properties \cite{gatti_transforming_2007, ferretti_bridging_2014, vanzini_dynamical_2017}. Despite the absence of a rigorous relation between KS spectra and experimental band structures, the qualitative agreement between the two (upon band gap opening) suggests the existence of a deeper physical connection and pushed some to consider KS eigenvalues as an approximation to ionization energies \cite{chong_interpretation_2002, stowasser_what_1999, gritsenko_physical_2003}.

In Hartree-Fock (HF) theory this connection is more straightforward. In fact, in the absence of electronic relaxations, Koopmans' theorem states that the expectation values of the HF Hamiltonian over the single empty or filled spin-orbitals are equal to the electron addition or removal energies \cite{koopmans_uber_1934}. Still, two major drawbacks remain: (i) the lack of electronic correlations in HF, and (ii) accounting for electronic relaxations invalidates this equivalence between expectation values and electron addition/removal energies, and turns the total HF energy into a non-linear concave function of the particle number \cite{li_piecewise_2017} accompanied by a systematic overestimation of energy gaps in molecules and solids.

Meanwhile, in practical KS-DFT calculations the eigenvalues suffer also from the approximations to the unknown exchange and correlation energy functional. In local and semi-local approximations the IP -- which in principle should be exact -- is systematically underestimated when taken as the opposite of $\varepsilon_{\rm HO}$ because of the erroneous convex behaviour of most functionals \cite{li_piecewise_2017, mori-sanchez_localization_2008} which deviate from the exact piecewise-linearity (PWL) of the total energy as a function of the number of electrons \cite{perdew_density-functional_1982}. The convexity of the energy curve leads to a discrepancy between the finite and differential energy differences -- with the former corresponding to $E(N-1) - E(N)$ and the latter to $\varepsilon_{\rm HO}$ \cite{janak_proof_1978, perdew_density-functional_1982, stein_curvature_2012, kraisler_piecewise_2013}. This discrepancy is often associated with the ``self-interaction" or ``delocalization" error present in (semi-)local DFT functionals \cite{cococcioni_linear_2005, kulik_density_2006, ruzsinszky_density_2007, cohen_insights_2008, mori-sanchez_localization_2008}. While the missing PWL and the self-interaction error are related \cite{mori-sanchez_many-electron_2006}, they might not be equivalent \cite{kronik_piecewise_2020}; the Koopmans integral (KI) functional \cite{borghi_koopmans-compliant_2014}, described in detail below, restores PWL (with an orbital-density dependence) while not improving the self-interaction or delocalization error of the base functional.

Many efforts have been made to go beyond the limitations of DFT and improve the agreement with experimental observations or have direct access to more physical features such as excited-state properties. While diagrammatic approaches, like $GW$ \cite{aryasetiawan_thegwmethod_1998, reining_gw_2018} and other more refined many-body perturbation theory methods \cite{shishkin_accurate_2007, shishkin_self-consistent_2007, chen_accurate_2015} provide a state-of-the-art baseline, functional approaches such as  Koopmans-compliant functionals \cite{di_valentin_piecewise_2014, borghi_koopmans-compliant_2014, nguyen_koopmans-compliant_2018, colonna_koopmans-compliant_2019, ma_using_2016} and range-separated or dielectric-dependent hybrid functionals \cite{stein_fundamental_2010, miceli_nonempirical_2018, chen_nonempirical_2018, bischoff_adjustable_2019, lorke_koopmans-compliant_2020, skone_nonempirical_2016} can show remarkable accuracy at significantly reduced computational costs.

The concept of piecewise-linearity of the total energy is the fundamental ingredient for the approach discussed in this work, a class of orbital-density-dependent (ODD) functionals that we termed as Koopmans-compliant (KC) \cite{dabo_non-koopmans_2009, dabo_koopmans_2010, di_valentin_piecewise_2014, borghi_koopmans-compliant_2014, ferretti_bridging_2014, nguyen_first-principles_2015, nguyen_first-principles_2016, colonna_screening_2018, nguyen_koopmans-compliant_2018, colonna_koopmans-compliant_2019}. These are energy functionals that satisfy a generalized PWL condition for every orbital, enforcing the following Koopmans-compliance condition \cite{dabo_koopmans_2010}
\begin{equation}
    \langle \phi_i | \hat{h}^{\rm KC} | \phi_i \rangle = \frac{dE^{\rm KC}}{d f_i} = \mathrm{constant} .
    \label{eq:kc-cond}
\end{equation}
Here $f_i$ is the occupation number of the $i$-th orbital $\phi_i$ and $\hat{h}^{\rm KC}$ the Koopmans-compliant Hamiltonian. The connection between the derivatives of the total energy with respect to the orbitals occupation and the expectation values of the Hamiltonian is ensured by Janak's theorem \cite{janak_proof_1978}. The condition above takes the exact concept of a PWL energy as a function of the total number of particles (i.e.\ the occupation of the HO orbital) and heuristically extends it to all the orbitals of the system, implementing a generalized version of Koopmans' theorem \cite{dabo_koopmans_2010}.

The constant in Eq.~\eqref{eq:kc-cond} -- let us call it $\lambda_i$ -- corresponds to the energy difference $\Delta E_i$ between the $N$-particle ground state and the relaxed $N \pm 1$-particle system where an electron has been removed or added to the $i$-th orbital, provided that the functional is piecewise linear. The generalized PWL condition discussed above, together with a correct description of the screening and relaxation effects due to the addition/removal of an electron, has been proven to be key for an accurate prediction of ionization potentials \cite{colonna_koopmans-compliant_2019} and electron affinities \cite{nguyen_first-principles_2015, nguyen_first-principles_2016, de_almeida_electronic_2021} of molecules. In extended periodic systems the situation is more complex and the localization of orbitals turns out to be fundamental in order to have effective KC corrections \cite{nguyen_koopmans-compliant_2018}; this feature comes out naturally from the energy minimization and ensures a seamless transition to the thermodynamic limit. Furthermore, the Wannier-like character of the resulting minimizing orbitals (so called ``variational'' orbitals) \cite{nguyen_koopmans-compliant_2018} provides a formal justification to related approaches where Koopmans-like corrections are directly applied to Wannier functions (WFs)~\cite{ma_using_2016, weng_wannier_2017, weng_wannierkoopmans_2020, stengel_self-interaction_2008, anisimov_transition_2005, anisimov_orbital_2007, wing_band_2021}.

That said, the localized and non-periodic nature of variational orbitals in Koopmans calculations makes it natural to use a supercell setup, even when perfectly periodic systems are considered -- apparently breaking the translation symmetries of the system. As a consequence, the single-particle eigenenergies obtained in supercell ($\Gamma$-sampling) calculations cannot be unfolded straightforwardly, since the existence of a band structure picture is linked to the compliance with Bloch's theorem. So, earlier works only focused on band gaps rather than on full band structures \cite{nguyen_koopmans-compliant_2018}, or rather applied KC-like corrections in a non-self-consistent manner directly to Kohn-Sham states \cite{ma_using_2016}. In the present work we show how to recover the full band structure description of the quasi-particle energies when working in an ODD framework. We show that when the minimizing orbitals are Wannier functions the orbital-density-dependent potentials inherit the Wannier translational property, and that this plays a fundamental role for the fulfillment of Bloch's theorem in KC functionals as well as in any other orbital-density-dependent approach like, e.g., the Perdew-Zunger (PZ) self-interaction correction \cite{perdew_self-interaction_1981, klupfel_effect_2012, lehtola_variational_2014}. 

In the following, we provide in Sec.~\ref{sec:kc-functionals} a general introduction of the framework of Koopmans-compliant functionals; in Sec.~\ref{sec:variational-orb} we highlight the importance of localization when performing KC calculations in extended systems; in Sec.~\ref{sec:bloch-theorem} we discuss the validity of Bloch's theorem within the framework of ODD functionals and in Sec.~\ref{sec:unfold-interpolate} we explain the method used to reconstruct the band structure from supercell $\Gamma$-only calculations. Finally, in Secs.~\ref{sec:comput-details}-\ref{sec:results} we discuss the computational details and the results obtained for a set of selected semiconductors and insulators.

\section{Methodology}

\subsection{\label{sec:kc-functionals}Koopmans spectral functionals}

Any KC construction starts with a base density functional, $E^{\rm DFT}[\rho]$, to which an orbital-density-dependent correction $\Pi^{\rm KC}_i[\rho,\rho_i]$ is applied:
\begin{equation}
    E^{\rm KC}[\{\rho_i\}] = E^{\rm DFT}[\rho] + \sum_i \Pi_i^{\rm KC}[\rho,\rho_i] ,
    \label{eq:kc-functionals}
\end{equation}
where $\rho(\br) = \sum_i f_i |\phi_i(\br)|^2$ is the total electronic density and $\rho_i(\br) = f_i|\phi_i(\br)|^2$ is the density of the $i$-th orbital with occupation $f_i$. By requiring $E^{\rm KC}$ to be compliant with Eq.\eqref{eq:kc-cond} one obtains the expression for the derivative of the $\Pi^{\rm KC}_i$ terms:
\begin{equation}
    \left.\frac{d\Pi_i^{\rm KC}}{df_i} \right|_{f_i=s} = - \left. \frac{dE^{\rm DFT}}{df_i} \right|_{f_i=s} + \lambda_i
    \label{eq:derivative-pi-occupation}
\end{equation}
where $s$ can take any value between 0 and 1. Upon integration between 0 and a generic occupation number $f_i \in [0,1]$, gives
\begin{equation}
    \Pi^{\rm KC}_i = - \int_{0}^{f_i} \braket{\phi_i | \hat{h}^{\rm DFT}(s) | \phi_i} ds + f_i \lambda_i ,
    \label{eq:ki-correction-relaxed}
\end{equation}
where $\hat{h}^{\rm DFT}$ represents the KS Hamiltonian and Janak's theorem \cite{janak_proof_1978} has been used,
\begin{equation}
    \left. \frac{dE^{\rm DFT}}{df_i} \right|_{f_i=s}=\braket{\phi_i | \hat{h}^{\rm DFT}(s) | \phi_i} .
    \label{eq:janak}
\end{equation}
As it can be seen from Eq.~\eqref{eq:ki-correction-relaxed}, $\Pi^{\rm KC}_i$ is made of two terms: the first removes orbital-by-orbital the non-linear dependence of the energy with respect to occupation $f_i$, while the second adds back a term that is linear with respect to $f_i$, with a slope $\lambda_i$ that can be chosen in a number of ways. Different choices for $\lambda_i$ give rise to different KC functionals \cite{borghi_koopmans-compliant_2014}. In the KI flavour, for instance, $\lambda_i$ is given by DFT total energy difference: for occupied states $\lambda_i = E^{N} - E^{N-1}_i$, with $E^{N-1}_i$ being the energy of the system where the occupied orbital $i$ is emptied, while for empty states $\lambda_i = E^{N+1}_i - E^{N}$, with $E^{N+1}_i$ being the energy of the system where the unoccupied orbital $i$ is filled. In both cases, the emptied or filled orbital is kept frozen while all the other orbitals are allowed to relax. Notably, for KI the base functional remains unchanged at integer occupation numbers ($f_i=0$ and $f_i=1$), but the derivatives (hence the expectation values of the Hamiltonian) differ \cite{borghi_koopmans-compliant_2014}. The KI functional can be expected to give results similar to those coming from $\Delta$SCF calculations performed with semi-local functionals when applied to small molecules (see later discussion for the solid-state limit).

As mentioned before, when evaluating the terms in Eq.~\eqref{eq:ki-correction-relaxed} one needs to consider that any modification in the occupancy of a single-particle density has an effect on all the other orbitals. Our strategy in this situation is to neglect the orbitals relaxation during the construction of the Koopmans corrective terms $\Pi^{\rm KC}_i$, and account for it \emph{a posteriori} by scaling the unrelaxed correction terms via scalar, orbital-dependent screening parameters $\alpha_i$. So, Eq.~\eqref{eq:kc-functionals} takes the form
\begin{equation}
    E^{\rm KC}[\{ \rho_i \}] = E^{\rm DFT}[\rho] + \sum_i \alpha_i \Pi^{\rm uKC}_i[\rho,\rho_i] ,
    \label{eq:kc-functionals-2}
\end{equation}
and $\Pi^{\rm uKC}_i$ represents the unrelaxed KC correction, i.e. when the orbitals are not allowed to relax after the addition/removal of an electron. In the KI case this is simply
\begin{equation}
\begin{split}
    \Pi^{\rm uKI}_i[\rho,\rho_i] = & \left\{ E^{\rm DFT}[\rho-\rho_i] - E^{\rm DFT}[\rho] \right\} \\
    + f_i & \left\{ E^{\rm DFT}[\rho-\rho_i+n_i] - E^{\rm DFT}[\rho-\rho_i] \right\} ,
\end{split}
    \label{eq:ki-correction-unrelaxed}
\end{equation}
where $n_i(\br) = \rho_i^{f_i=1}(\br) = |\phi_i(\br)|^2$ . It is apparent from this expression that $\Pi^{\rm uKI}_i$ is zero for $f_i=0$ or $f_i=1$. One can also apply the KI linearization to the orbital-dependent self-interaction-corrected Perdew-Zunger (PZ) functional \cite{perdew_self-interaction_1981}, leading to the so-called KIPZ functional \cite{borghi_koopmans-compliant_2014}. This is more computationally expensive than the simple KI, but it has the advantage of being exactly one-electron self-interaction-free. Detailed expressions for KI, KIPZ and their derivatives are given in Appendix~\ref{app:kc-functionals-potentials}. As an aside, we also note that the derivatives of $\Pi^{\rm uKI}_i$ and of $\Pi^{\rm uKIPZ}_i$ with respect to $f_j$ for $j \neq i$ vanish, which justifies \emph{post-hoc} the absence of cross-derivatives when going from Eq.~\eqref{eq:kc-functionals} to Eq.~\eqref{eq:derivative-pi-occupation}. The effect of the orbitals relaxation is added by renormalizing each $\Pi^{\rm KC}_i$ via a screening factor $\alpha_i$. While the screening should generally be accounted for via non-local functions in space (see Ref.~\cite{colonna_screening_2018}), in KC functionals we consider an approximated, yet effective, scalar form that can be calculated fully from first-principles either with a linear response approach \cite{colonna_screening_2018} -- averaging the microscopic dielectric function -- or via finite differences \cite{nguyen_koopmans-compliant_2018}. In this work we followed the second strategy, where each screening parameter $\alpha_i$ optimizes the matching between the right derivative of the energy when the orbital is empty, $\lambda_i(f_i=0)$, and the left derivative of the energy when the orbital is completely filled, $\lambda_i(f_i=1)$. In the following we report the expression used for the screening parameters while for a more detailed discussion we refer to \cite{nguyen_koopmans-compliant_2018, dabo_koopmans_2010}:
\begin{equation}
    \alpha_i = \alpha_i^{(0)} \frac{\Delta E_i - \braket{\phi_i|\hat{h}^{\rm DFT}|\phi_i}}{\lambda_i^{\alpha^{(0)}} - \braket{\phi_i|\hat{h}^{\rm DFT}|\phi_i}} ,
    \label{eq:alpha}
\end{equation}
where $\Delta E_i = E^{\rm KC}(f_i=1) - E^{\rm KC}(f_i=0)$, $\alpha_i^{(0)}$ is a trial value for the screening parameter and, by referring to Eq.~\eqref{eq:kc-cond}$, \lambda_i^{\alpha^{(0)}}$ is the expectation value over the $i$-th orbital of the Koopmans Hamiltonian where the screening parameters have been set to the trial values $\{ \alpha_i^{(0)} \}$.

As is the case for the PZ functional, the orbital-dependent nature of KC functionals breaks the invariance of the total energy with respect to unitary transformations of the manifold of occupied orbitals. Consequently, the ground-state energy cannot be found via a self-consistent diagonalization, but instead requires a direct minimization of the functional
\begin{equation}
    E^{\rm KC}_{\rm GS} = \min_{\hat{\rho}} \min_{\{ \phi_i \}_{\hat{\rho}}}
    \left\{ E^{\rm KC}[\{ \rho_i \}] - \sum_{ij} \Lambda_{ij} \left( \braket{\phi_i | \phi_j} -\delta_{ij} \right) \right\} ,
    \label{eq:kc-minimum}
\end{equation}
where we have included the usual orthonormality constraint on the single-particle wavefunctions. Minimizing this functional typically involves (i) an outer-loop that searches for the optimal manifold of occupied orbitals $\hat{\rho}=\sum_i \ket{\phi_i} \bra{\phi_i}$, and (ii) an inner-loop that, for a fixed manifold $\hat{\rho}$, determines the set of variational orbitals, that minimizes the energy \eqref{eq:kc-functionals-2} with respect to all possible unitary rotations \cite{goedecker_critical_1997, borghi_variational_2015, lehtola_variational_2014}.

At its minimum, the functional satisfies the Pederson condition \cite{pederson_localdensity_1984, pederson_densityfunctional_1985}:
\begin{equation}
    \braket{\phi_i | \hat{h}^{\rm KC}[\rho_j] | \phi_j} = \braket{\phi_i | \hat{h}^{\rm KC}[\rho_i] | \phi_j} ,
    \label{eq:pederson-condition}
\end{equation}
which means that the matrix of Lagrangian multipliers becomes Hermitian, providing the matrix representation of the KC Hamiltonian on the variational orbitals. The expression for the action of the KC Hamiltonian on the variational orbitals is then defined by the gradient of the energy functional (for more details see Appendix~\ref{app:kc-functionals-potentials}):
\begin{equation}
    \frac{\delta E^{\rm KC}}{\delta \rho_i} = \hat{h}^{\rm KC}[\rho,\rho_i] \equiv \hat{h}^{\rm DFT}[\rho] + \alpha_i \hat{v}^{\rm KC}[\rho,\rho_i] .
    \label{eq:gradient}
\end{equation}
While the authors of Ref.~\cite{vydrov_tests_2007} suggested to interpret the diagonal elements of the Lagrangian multipliers matrix as excitation energies, here we follow the prescription of Ref.~\cite{stengel_self-interaction_2008} and consider the canonical diagonal representation of the matrix $\Lambda_{ij}$ to interpret its eigenvalues as quasi-particle energies (and its eigenvectors as ``canonical'' orbitals, as opposed to the variational ones that minimize the functional). 

Before moving on and facing more specifically the problem of periodic systems, we mention here a current limitation of the approach, due to the impossibility of treating a generic density matrix. KC functionals provide a correction only for the diagonal elements of the occupation matrix, whereas no correction is applied to off-diagonal occupations $f_{ij} = \braket{\phi_i | \hat{\rho} | \phi_j}$, which forces us to have a density matrix of the form $\hat{\rho} = \sum_i f_i \ket{\phi_i} \bra{\phi_i}$. At zero temperature, for a system with a non-zero band gap the occupation matrix is trivially block-diagonal with the identity on the occupied block and the null matrix on the empty block. The occupation matrix of a metal, on the other hand, does not have this structure and unitary transformations of the orbitals will mix the occupied and empty manifolds and lead to some non-zero off-diagonal elements -- i.e.\ to a density matrix of the form $\hat{\rho} = \sum_{ij} f_{ij} \ket{\phi_i} \bra{\phi_j}$. As a consequence, only systems possessing a finite band gap at the DFT level are accessible by this approach, whereas metallic systems are, for the moment, not treated.

\subsection{\label{sec:variational-orb}Localization: Wannier functions as variational orbitals}

The breaking of unitary invariance in Koopmans functionals gives rise to a unique set of minimizing, or variational, orbitals. While in DFT any unitary rotation of the KS states does not modify the total energy, in KC functionals such a transformation of the variational orbitals alters the total energy, despite the fact that the density is left unchanged. We stress that while this is true in general, KI at integer occupation numbers represents an exception as it is also invariant with respect to unitary transformations; although not changing the KI total energy, a unitary transformation does modify its derivatives with respect to the orbital occupations, and thus the KI spectrum. It is therefore necessary to remove this ambiguity by defining KI as the limit of KIPZ with the PZ contribution going to zero~\cite{borghi_koopmans-compliant_2014} (see also Appendix~\ref{app:kc-functionals-potentials}).

Typically the KC variational orbitals tend to be localized in space; this is because the minimization of a KC functional for the occupied states is dominated by the gradient of the PZ term (see Appendix~\ref{app:kc-functionals-potentials}). The very fact that these orbitals are localized is a characteristic feature of KC functionals and key to obtain non-vanishing KC corrections in extended systems \cite{nguyen_koopmans-compliant_2018}. If we instead consider the canonical KS orbitals coming from local or semi-local density-functionals, those are typically delocalized and variations of energy due to a change in the occupancy -- as calculated via a $\Delta$SCF calculation -- recover the value of the KS eigenenergies, that are a poor approximation for excitation energies \cite{perdew_understanding_2017, kraisler_fundamental_2014, nguyen_koopmans-compliant_2018}. Analogously, if we were to apply the Koopmans correction of Eq.~\eqref{eq:ki-correction-unrelaxed} to a fully delocalized state this correction would be vanishingly small and Koopmans functionals would be completely ineffective.

In addition to their localized character the variational orbitals in periodic systems typically possess the translation property of Wannier functions \cite{nguyen_koopmans-compliant_2018}:
\begin{equation}
    w_{\bR}(\br - \bR') = w_{\bR + \bR'}(\br),
    \label{eq:trans-prop-wannier}
\end{equation}
where $\bR$ and $\bR'$ represent any pair of Bravais lattice vectors. We therefore say that the variational orbitals of KC functionals in extended, periodic systems are ``Wannier-like", namely they constitute an orthonormal set of functions that (i) are localized in space and (ii) satisfy Eq.~\eqref{eq:trans-prop-wannier}, i.e. they can be labeled with a corresponding lattice vector. Indeed, in periodic systems the KIPZ variational orbitals typically resemble maximally localized Wannier functions (MLWFs) \cite{marzari_maximally_1997}, providing also a justification for the non-self consistent application of KI corrections on top of MLWFs~\cite{ma_using_2016}. As mentioned earlier in this section, because of the unitary invariance of the the KI energy, the variational orbitals must be obtained by the minimization of the KIPZ energy with an arbitrarily small PZ term. This is equivalent to minimize the KIPZ energy while constraining the density to match the ground-state density of the underlying DFT functional. 
For this reason, MLWFs obtained from the KS states represent a very good choice for the KI variational orbitals, while for KIPZ they constitute a very good initial guess for the minimization.

While the strong localization of variational orbitals forces us to resort to a supercell approach and seems to break the translation symmetries of the system, the Wannier-like character turns out to be the key to prove the periodicity of the KC potential over the primitive cell, legitimizing the reconstruction of energy dispersions within the Brillouin zone of the primitive cell of the system via an unfolding procedure.

\subsection{\label{sec:bloch-theorem}Bloch's theorem}

In periodic systems, the emergence of the crystal momentum $\bk$ as a quantum number is a natural consequence of the fact that the effective crystal Hamiltonian commutes with all the translation operators, $\{ \hat{T}_{\bR} \}$ ($\bR$ are the Bravais lattice vectors of the primitive cell). As a consequence of Bloch's theorem, and more generally of group theory, the irreducible representations (labeled by $\bk$) of the translation group allow for a block-diagonal representation of the Hamiltonian and, consequently, for a band structure description of the energy spectrum. We will call such a Hamiltonian ``Bloch-compliant".

Before discussing Bloch's theorem in the context of KC functionals, let us first consider what happens in standard DFT, where the energy functional and the Hamiltonian depend only on the total density. If we consider a perfectly periodic solid in a supercell calculation with a $\Gamma$-point sampling of the Brillouin zone, the periodicity of the density over the primitive cell is not imposed \emph{a priori}, but it typically emerges during the energy minimization (we exclude exotic ground states, like those with charge density waves, where the periodicity of the density is not commensurate with that of the lattice). The periodicity of the density is then inherited by the Hamiltonian, which is therefore Bloch-compliant. The KS orbitals, meanwhile, will be exactly periodic with the boundary conditions of the supercell but they might not have a Bloch-like form with respect to the primitive cell. Contrast this with a primitive cell calculation, where the validity of Bloch's theorem is assumed since the beginning: the KS orbitals are forced to be Bloch functions and the density and the operators are periodic over the primitive cell. As a consequence, the KS potential is periodic over the primitive cell -- i.e. the Hamiltonian commutes with the translation operators -- and the eigenstates are labeled with corresponding $\bk$-vectors.

At odds with DFT, where the total density is the only quantity entering the Hamiltonian, KC functionals (and their Hamiltonians) depend on the set of variational orbital densities, and therefore the periodicity (over the primitive cell) of the total density alone is not sufficient to obtain a periodic potential. In this case a more stringent condition is needed and in the following we show that this extra condition is given by the Wannier-like character of the variational orbitals. If the variational orbitals satisfy Eq.~\eqref{eq:trans-prop-wannier}, the KC potential turns out to have the periodicity of the primitive cell and, therefore, the KC Hamiltonian fulfills the hypothesis of Bloch's theorem.

At the minimum of the KC energy functional the matrix of Lagrangian multipliers ensuring the orthonormality of the variational orbitals is Hermitian \cite{pederson_densityfunctional_1985, stengel_self-interaction_2008}, and the KC Hamiltonian is defined as
\begin{equation}
\begin{split}
    \hat{h}^{{\rm KC}} &= \hat{h}^{{\rm DFT}} + \hat{v}^{{\rm KC}} \\
    &= \sum_i \left( \hat{h}^{\rm DFT}[\rho] + \hat{v}^{\rm KC}[\rho,\rho_i] \right) \ket{\phi_i} \bra{\phi_i}
    \end{split}
    \label{eq:kc-hamiltonian}
\end{equation}
where $\{ \phi_i \}$ are the variational orbitals and $\{ \hat{v}^{\rm KC}[\rho,\rho_i] \}$ are the ODD potentials introduced in Sec.~\ref{sec:kc-functionals} and explicitly shown in Appendix~\ref{app:kc-functionals-potentials}. Because the ODD potentials $\{ \hat{v}^{\rm KC}[\rho,\rho_i] \}$ are built from the variational orbitals and thus are periodic only in the supercell (and not in the primitive cell), each $\hat{v}^{\rm KC}[\rho,\rho_i]$ individually breaks the translation symmetry of the system. However, it can be shown that, if the variational orbitals are Wannier functions (i.e.\ they satisfy Eq.~\eqref{eq:trans-prop-wannier}) the potential $\hat{v}^{\rm KC} = \sum_i \hat{v}^{\rm KC}[\rho,\rho_i] \ket{\phi_i} \bra{\phi_i}$ is periodic over the primitive cell. While this result is general and applies to any ODD Hamiltonian, for simplicity here we give the proof for the PZ Hamiltonian where the DFT Hamiltonian is augmented by the PZ potential $\hat{v}^{\rm PZ} = - \sum_i \hat{v}^{\rm Hxc}[\rho_i]\ket{\phi_i}\bra{\phi_i}$ (where $\{ \phi_i \}$ is the set of PZ variational orbitals).

From now on we assume the Wannier-like character of the variational orbitals -- which also implies the periodicity of the total density over the primitive cell -- and we replace the generic orbital index $i$ with a composite index $(\bR,n)$ labeling the Wannier functions in terms of the primitive cell index $\bR$ and a band index $n$ \cite{marzari_maximally_2012}. Without loss of generality and in order to keep the notation simple we consider only one band and in the following we drop the band index $n$, so $v^{\rm Hxc}([\rho_i],\br) = v^{\rm Hxc}([\rho_{\bR}],\br)$, and the PZ potential becomes:
\begin{equation}
    \hat{v}^{\rm PZ} = - \sum_{\bR} \hat{v}^{Hxc}[\rho_{\bR}] \ket{w_{\bR}} \bra{w_{\bR}} .
    \label{eq:pz-potential}
\end{equation}

As shown in Appendix~\ref{app:wannier-occupations}, the occupation numbers of the Wannier functions are independent of the lattice vector $\bR$: $f_{\bR} = f_{\bm 0}$. By definition, the Wannier orbital densities $\rho_{\bR}(\br) = f_{\bm 0} |w_{\bR}(\br)|^2$ fulfill the same translation property \eqref{eq:trans-prop-wannier} as $w_{\bR}(\br)$. The crucial step is to show that also the ODD potentials $\hat{v}^{\rm Hxc}[\rho_{\bR}]$ satisfy the same property. Let us consider the Hartree ODD potentials first:
\begin{equation}
\begin{split}
    v^{\rm H}([\rho_{\bR}],\br - \bR') & = \int_V d\br' \frac{\rho_{\bR}(\br')}{|\br - \bR' - \br'|} \\ 
    & = \int_V d\br' \frac{\rho_{\bR}(\br' - \bR')}{|\br - \br'|} \\
    & = \int_V d\br' \frac{\rho_{\bR + \bR'}(\br')}{|\br - \br'|} \\
    & = v^{\rm H}([\rho_{\bR + \bR'}],\br) ,
\end{split}
\end{equation}
where $V$ is the supercell volume and we made use of property \eqref{eq:trans-prop-wannier} on $\rho_{\bR}$. The situation is even simpler for the xc ODD potentials $\hat{v}^{\rm xc}[\rho_{\bR}]$; at LDA and GGA levels, the xc functional is $E_{\rm xc}[\rho] = \int d\br f(\rho(\br), \nabla\rho(\br))$, hence the xc potential is given by $v^{\rm xc}([\rho_{\bR}],\br) = \frac{df}{d\rho_{\bR}}(\rho_{\bR}(\br), \nabla\rho_{\bR}(\br))$, and it is straightforward to show that $v^{\rm xc}([\rho_{\bR}],\br-\bR') = v^{\rm xc}([\rho_{\bR+\bR'}],\br)$.

Overall then we obtain
\begin{equation}
    v^{\rm Hxc}([\rho_{\bR}],\br - \bR') = v^{\rm Hxc}([\rho_{\bR + \bR'}],\br)  \qquad \forall \bR, \bR' ,
    \label{eq:trans-prop-pot}
\end{equation}
which means that the potentials $\{ -\hat{v}^{\rm Hxc}[\rho_{\bR}] \}$ satisfy the same translation properties as the set of WFs $\{ w_{\bR} \}$. Using Eqs.~\eqref{eq:trans-prop-wannier} and \eqref{eq:trans-prop-pot} it is straightforward to show that the PZ potential $v^{\rm PZ}(\br) = \braket{\br | \hat{v}^{\rm PZ} | \br}$ is periodic over the primitive cell (an alternative and equivalent proof showing that the PZ potential commutes with the translation operators $\{\hat{T}_{\bR}\}$ is given in Appendix~\ref{app:bloch-theorem}):
\begin{equation}
    \begin{split}
    v^{\rm PZ}(\br+\bR) &= - \sum_{\bR'} v^{\rm Hxc}([\rho_{\bR'}],\br+\bR) w_{\bR'}(\br+\bR) w^*_{\bR'}(\br+\bR) \\
    &= - \sum_{\bR'} v^{\rm Hxc}([\rho_{\bR'-\bR}],\br) w_{\bR'-\bR}(\br) w^*_{\bR'-\bR}(\br) \qquad (\bR'-\bR \rightarrow \bR') \\
    &= - \sum_{\bR'} \hat{v}^{\rm Hxc}[\rho_{\bR'}] w_{\bR'}(\br) w^*_{\bR'}(\br) \\
    &= v^{\rm PZ}(\br) .
    \end{split}
    \label{eq:periodic-pot-pz}
\end{equation}
The remainder of the PZ Hamiltonian -- the DFT Hamiltonian -- commutes already with all the translation operators, thus it fulfills Bloch's theorem. The Wannier-like property of the orbital-dependent potentials $\hat{v}^{\rm PZ}[\rho,\rho_{\bR}]$, Eq.~\eqref{eq:trans-prop-pot}, and the consequent periodicity of the PZ potential $\hat{v}^{\rm PZ}$, Eq.~\eqref{eq:periodic-pot-pz}, are the central results of this paper. They show that, under the assumption of the Wannier-like character of the variational orbitals, the PZ potential has the periodicity of the primitive cell and therefore it satisfies the hypothesis of Bloch's theorem. 

Although the KC Hamiltonian is more complex than the PZ one discussed here, it is made of orbital-density-dependent potentials of the very same nature of the PZ ones, alongside scalar terms -- meaning that they do not depend on $\br$ -- or terms that depend only on the total density $\rho$. The latter are trivially periodic on the primitive cell because of the periodicity of the total density. The proof given above thus readily applies also to the KC potential and so
\begin{equation}
    v^{\rm KC}(\br+\bR) = v^{\rm KC}(\br) \qquad \forall \bR \ .
    \label{eq:periodic-pot-kc}
\end{equation}

For further evidence of the compliance of the KC Hamiltonian with Bloch symmetries, we can consider the unitary transformation between Wannier functions and Bloch functions that acts as a Fourier transform from $\bR$-space to $\bk$-space:
\begin{equation}
    h^{\rm KC}(\bk,\bk') = \sum_{\bR,\bR'} e^{-i\bk \cdot \bR} e^{i\bk' \cdot \bR'} \braket{w_{\bR} | \hat{h}^{\rm KC}_{\bR'} | w_{\bR'}} .
    \label{eq:double-k-ham}
\end{equation}
Using Eq.~\eqref{eq:trans-prop-pot}, it can be easily shown that the expression above is proportional to $\delta_{\bk, \bk'}$ yielding a matrix that, in a Bloch-like representation, is block-diagonal over $\bk$.

To summarize, we have shown that when the variational orbitals are Wannier functions (i.e.\ they satisfy Eq.~\eqref{eq:trans-prop-wannier}) the potential $\hat{v}^{\rm KC}$ defined in Eq.~\eqref{eq:kc-hamiltonian} is periodic over the primitive cell, making the Koopmans Hamiltonian Bloch-compliant. As an aside, the Wannier-like nature of the orbitals and the Bloch-compliance of KC functionals also makes it possible to develop a primitive cell implementation of KC functionals, for direct access to the band structure without the need of supercell calculations and of an unfolding procedure; details about the implementation and the results are reported in Ref.~\cite{colonna_koopmans_2022}. As we already mentioned above, the assumption of having Wannier-like variational orbitals is justified by the observation that the minimization of KC and PZ functionals in extended systems leads to orbitals with these properties. While this has occurred in all the systems so far considered, it is important to remark that the lack of the Wannier-like character in the variational orbitals would prevent from applying the Bloch's theorem and implies the actual breaking of the translation symmetry of the system. Such situations are presumably as sporadic as when in standard DFT the periodicity of the ground-state density is not commensurate to that of the lattice potential and they should not be confused with special system-dependent gauge invariances that KC functionals might have. Despite the non-unitary invariance of the functionals, there is indeed no guarantee for the uniqueness of the set of variational orbitals and we cannot exclude \emph{a priori} that there might exist some gauges for which the translation symmetry of the lattice is broken.

Having proved the compliance of KC Hamiltonians with translation symmetries, in the following section we detail the unfolding procedure used to reconstruct the $\bk$-space Hamiltonian and find eigenvalues at any point of the Brillouin zone, starting from supercell $\Gamma$-sampling calculations.

\subsection{\label{sec:unfold-interpolate}Unfolding and interpolation of bands}

When simulating a bulk crystalline material, the infinite system is usually studied with Born-von Karman (BvK) boundary conditions, which introduces a discretization of the $\bk$-points inside the first Brillouin zone (BZ). Equivalently, one can study explicitly the BvK supercell containing the $N$ periodic replicas of the primitive cell, but in this case one no longer has direct access to the band structure of the primitive cell.
\begin{figure*}[htp]
    \centering
    \includegraphics[width=0.8\textwidth]{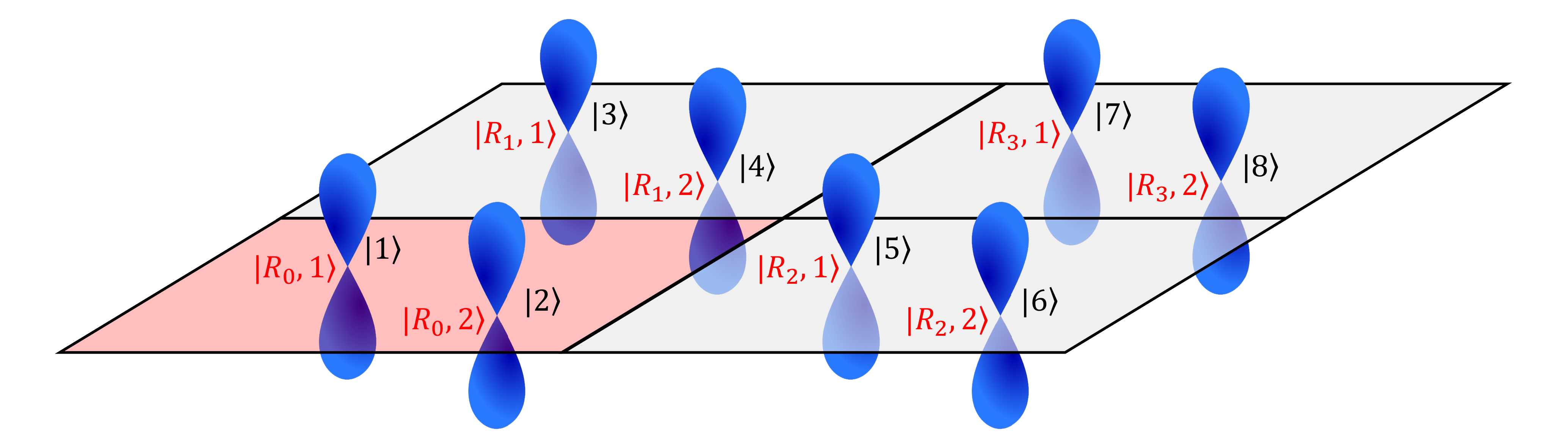}
    \caption{Schematic representation of a two-dimensional 2-band model showing the connection between the primitive and supercell Wannier representations. In the primitive picture with a $2\times 2$ sampling of the BZ, the WFs are identified with the pair of labels $\{\bR,n\}$ (red labels): the cell index $\bm{R}$ taking four values and the band index $n$ taking two values. In the $2\times 2$ supercell with $\Gamma$-sampling of the BZ, the eight WFs are labeled by only one quantum number (black labels), i.e.\ the supercell band index $\alpha$ running over the eight states.}
    \label{fig:pcell-vs-scell}
\end{figure*}
In order to recover this band structure, several methods have been developed \cite{boykin_practical_2005, lee_band_2005, ku_unfolding_2010, popescu_extracting_2012, huang_general_2014, medeiros_effects_2014, zheng_quantum_2015}; our approach follows the same strategy of \cite{lee_band_2005} and exploits the Wannier nature of the variational orbitals. The matrix elements of the $\bk$-space Hamiltonian are obtained from those given by the variational orbitals via a (double) Fourier transform as in Eq.~\eqref{eq:double-k-ham}. If the Wannier Hamiltonian is compliant with the translation symmetries of the system, the expression reduces to
\begin{equation}
    h^{\rm KC}_{mn}(\bk) = \braket{\psi_{m\bk} | \hat{h}^{\rm KC} | \psi_{n\bk}} = \sum_{\bR} e^{i \bk \cdot \bR} \braket{w_{\bm{0}m} | \hat{h}^{\rm KC}[\rho_{\bR n}] | w_{\bR n}} = \sum_{\bR} e^{i \bk \cdot \bR} h^{\rm KC}_{mn}(\bR) ,
    \label{eq:hk-unfold}
\end{equation}
where we have defined $h^{\rm KC}_{mn}(\bR) = \braket{w_{\bm{0}m} | \hat{h}^{\rm KC}[\rho_{\bR n}] | w_{\bR n}}$. The diagonalization of this matrix yields the energies $\varepsilon_{n\bk}$ at any $\bk$-point.

In the supercell approach the Brillouin zone reduces to a single point; as a consequence, the supercell Hamiltonian in the Wannier representation loses the information about the lattice vectors $\{\bR\}$ and its matrix elements are labeled by the supercell index only. In order to reconstruct the $\bk$-space Hamiltonian of Eq.~\eqref{eq:hk-unfold}, one must reconstruct the composite index $\{\bR,n\}$ of each WF from its supercell-picture index ${\alpha}$ (see Fig.~\ref{fig:pcell-vs-scell}). An effective way to do this is to first choose a reference primitive cell and define the orbitals with the centers inside it as the $\bR=\bm{0}$ Wannier functions. The second step consists of comparing all the other WFs in the supercell with those in the reference cell. If the Wannier translation property holds, we are able to connect each WF to its reference function $w_{\bm{0} n}$ and lattice vector $\bR$, defined as the distance between the centers of the two functions. If the system has more functions sharing the same center, one can look at the second-order moments ($\braket{x^2}$, $\braket{y^2}$, $\braket{z^2}$) to have a more detailed signature of WFs and, if needed, can move towards higher-order spatial moments until the character of each Wannier function is unequivocally defined \cite{shelley_automated_2011}.

As argued in Ref.~\cite{lee_band_2005}, Eq.~\eqref{eq:hk-unfold} not only applies to the points belonging to the $\bk$-mesh commensurate with the chosen supercell, but it is also an excellent interpolator. So, in order to calculate the band structure along any path in the Brillouin zone, we obtain the matrix elements of the $\bk$-space Hamiltonian by simply applying Eq.~\eqref{eq:hk-unfold} to any arbitrary $\bk$-point. In doing so, the matrix elements $h^{\rm KC}_{mn}(\bR)$, for any $\bR$-vector larger than the supercell are implicitly neglected; the accuracy of the approximation is higher the smaller the contribution from these terms, i.e.\ the more localized the variational orbitals are, or the larger the supercell becomes. A poorly interpolated band structure is usually a symptom of a significant contribution coming from the matrix elements corresponding to larger $\bR$-vectors, and increasing the size of the supercell improves the results.

\section{\label{sec:comput-details}Computational details}

All calculations are performed using the Quantum ESPRESSO (QE) distribution \cite{giannozzi_advanced_2017} and the Wannier90 code \cite{pizzi_wannier90_2020}. All the machinery of KC functionals has been implemented within a modified version of the Car-Parrinello (CP) code. In this work we use three different KC flavours: KI and KIPZ, described already in Sec.~\ref{sec:kc-functionals}, and a perturbative KIPZ (pKIPZ) approach where the orbitals are localized using the Marzari-Vanderbilt procedure \cite{marzari_maximally_1997} and the screening coefficients are obtained from a KI calculation, whereas the spectra are evaluated using the KIPZ Hamiltonian. The entire workflow has been automatized with a Python package based on the Atomic Simulation Environment (ASE) \cite{larsen_atomic_2017}.

A typical workflow for solids consists then of three main steps: (i) a standard DFT calculation is followed by a Wannierization of the KS states in order to obtain MLWFs that are either used as a non-self-consistent guess for the KI and pKIPZ variational orbitals (see also extended discussion in Appendix~\ref{app:kc-functionals-potentials}), or as initial guess for the minimization of the KIPZ functional; (ii) the calculation of the screening parameters and the application of the Koopmans correction; (iii) the unfolding procedure, described in Sec.~\ref{sec:unfold-interpolate}, to reconstruct the band structure in the BZ of the primitive cell.

The first part is fully driven by a $\bk$-point primitive cell calculation: the ground-state density is obtained via a self-consistent field calculation using the plane-wave (PWscf) QE implementation of DFT; a non-self-consistent calculation allows to obtain the KS valence and conduction states at any point on a regular Monkhorst-Pack mesh \cite{monkhorst_special_1976}. The Wannier90 code is then employed to obtain the rotation matrices $U^{(\bk)}$ defining the MLWFs for the occupied and empty manifolds separately. In all cases the base functional used is PBE and a $4\times4\times4$ Monkhorst-Pack mesh has been adopted. All the materials have been modeled using their stable experimental structure under standard conditions of temperature and pressure: cubic rock-salt for MgO and LiF, diamond (or zincblende) for all the rest. Experimental values for the lattice constants have been taken from Ref.~\cite{madelung_semiconductors_2004}. The electron-ion interactions have been modeled with the Schlipf-Gygi optimized norm-conserving Vanderbilt (ONCV) pseudopotentials (v1.2) \cite{schlipf_optimization_2015}. For the elemental compounds the energy cutoffs have been chosen from the convergence studies provided by the Standard Solid State Pseudopotentials (SSSP) precision protocol \cite{prandini_precision_2018}, while for the three binary compounds the cutoff was chosen by converging the KS-DFT band gap within $5\ \meV$.

In the second part, all the calculations are performed within a supercell (with $\Gamma$-point sampling) commensurate to the $\bk$-points grid used in the first part of the workflow. An interface written within the \texttt{pw2wannier90} code takes the Wannier matrices $U^{(\bk)}$ and the KS states to construct the MLWFs within the $\bR=\bm{0}$ primitive cell. The same code then shifts the set of orbitals $\{ w_{\bm{0} n} \}$ and obtains all the other WFs inside the supercell. Finally, when unfolding the Koopmans-compliant band structure, in case the $4\times4\times4$ sampling is not enough to obtain a good interpolation of the band structure, a smooth interpolation method (see Appendix~\ref{app:smooth-interpolation}) has been used to improve the quality of the interpolation.

In order to avoid ambiguity in the choice of MLWFs mixing different subspaces, we relied on the projected density-of-states to select the ``physically motivated'' initial projections. For covalent semiconductors like Si, C, Ge and GaAs, where the $s$ and $p$ orbitals contribute more or less equally in the energy range corresponding to the valence bands, the natural choice is that of $sp^3$ orbitals. In the case of GaAs and Ge, where $d$ semicore states are present, the Wannierization procedure gave rise to two sets of well distinguished groups of MLWFs ($d$-like and $sp^3$-bonding orbitals); however, for Ge we observed an unphysical mixing of the two types of orbitals. In this case, the two subsets of isolated valence bands were Wannierized separately in order to preserve the natural atomic character of the orbitals and to not mix Bloch functions corresponding to bands well separated in energy. This selection of the WFs follows also that of orbitalets used by Li \emph{et al.} \cite{li_localized_2018}, where the optimal orbitals are chosen via a localization procedure both in space and energy. A similar situation is observed in BN, MgO and LiF: $s$ and $p$ atomic orbitals contribute to separate ranges of energy and, for this reason, the hybridized choice $sp^3$ was considered ``less'' physical. In all cases, the KIPZ minimization reshapes the orbitals while maintaining the same atomic-like character.

For the evaluation of screening parameters $\{ \alpha_i \}$, we used the same strategy described in \cite{nguyen_koopmans-compliant_2018, ma_using_2016}. This involves constrained $N \pm 1$-electron calculations for each non-equivalent variational orbital, where the DFT or KIPZ energy is minimized while keeping an occupied (empty) variational orbital completely empty (filled). For equivalent orbitals here we intend orbitals that have the same screening parameter. This usually happens to orbitals having similar shapes (e.g. $sp^3$-like orbitals in silicon), where the differences due to the system's relaxation are negligible. In general, equivalent orbitals are selected when the spatial spread or the self-Hartree energy are the same within the chosen threshold. This approach avoids to perform redundant calculations and drastically reduces the overall computational cost.

To account for the spurious interactions between the charge and its periodic images, we employed Makov-Payne (MP) corrections for point-charge defects \cite{makov_periodic_1995}. However, for materials with small band gaps -- i.e.\ Si, Ge and GaAs -- MP corrections were not applied: the large dielectric constant of these materials ensures a natural screening of charged defects and makes the bare, uncorrected energies at $N \pm 1$ very close to their extrapolated values (for infinitely large supercells).

\section{\label{sec:results}Results}

\begin{figure*}[htp]
    \subfloat[Si: KI]{\includegraphics[width=.4\textwidth]{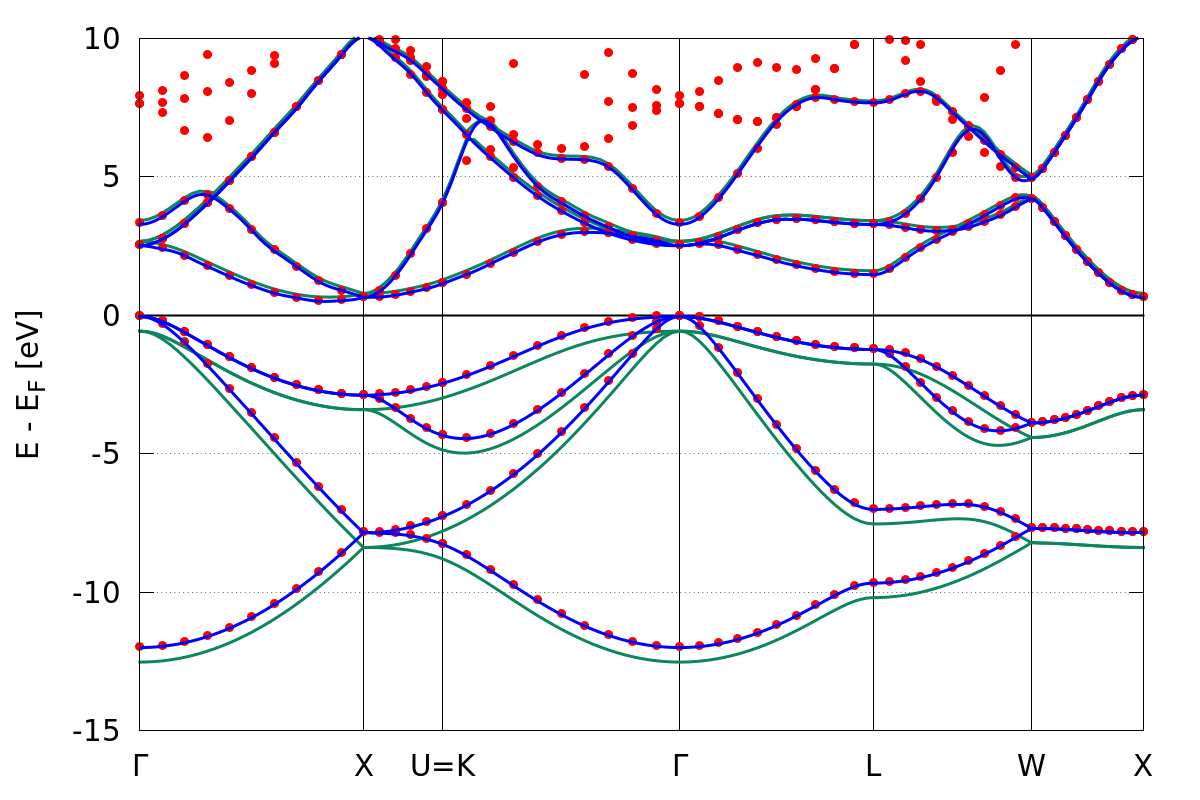}} \qquad
    \subfloat[Si: KIPZ]{\includegraphics[width=.4\textwidth]{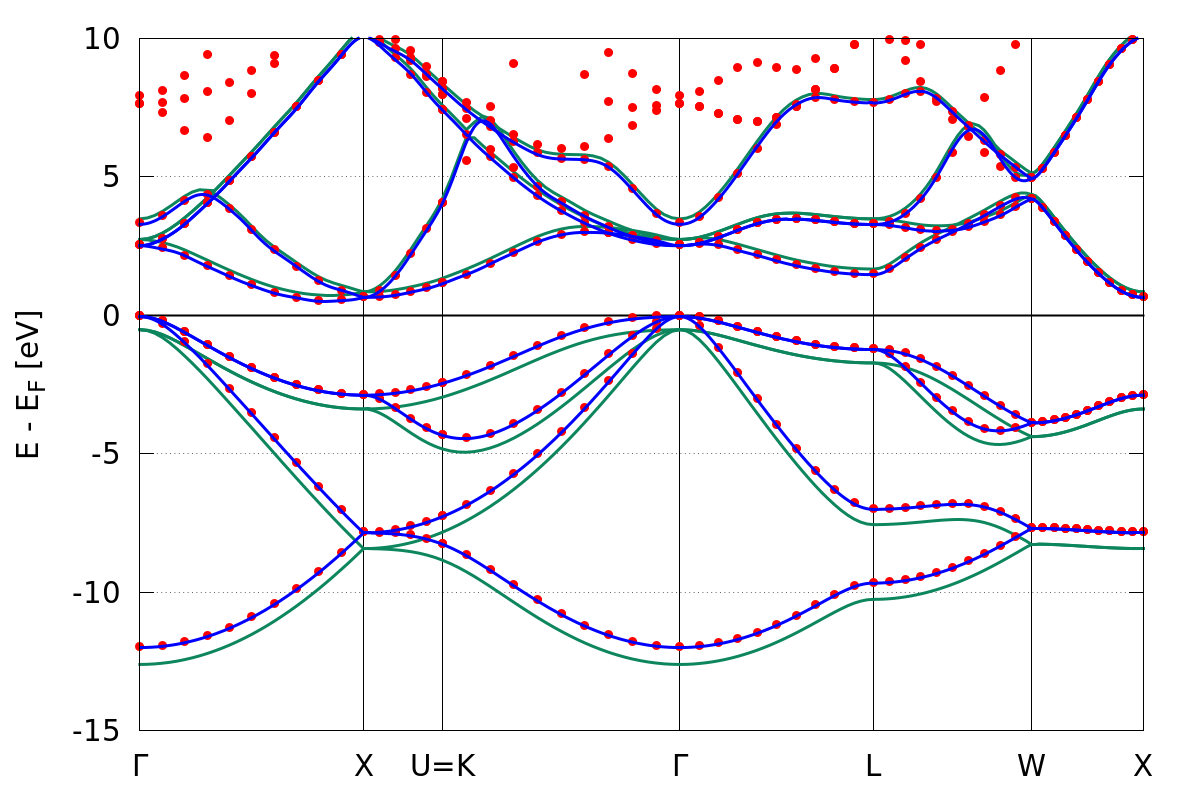}} \\
    \subfloat[C: KI]{\includegraphics[width=.4\textwidth]{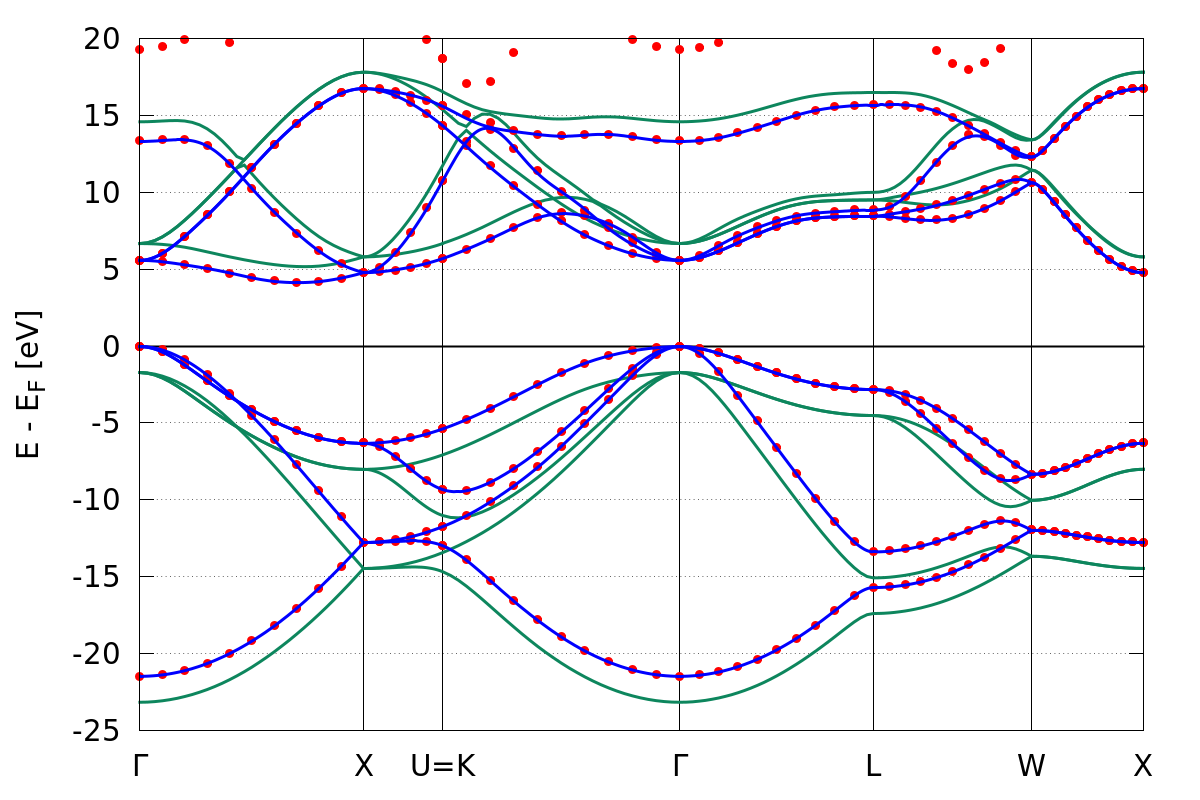}} \qquad
    \subfloat[C: KIPZ]{\includegraphics[width=.4\textwidth]{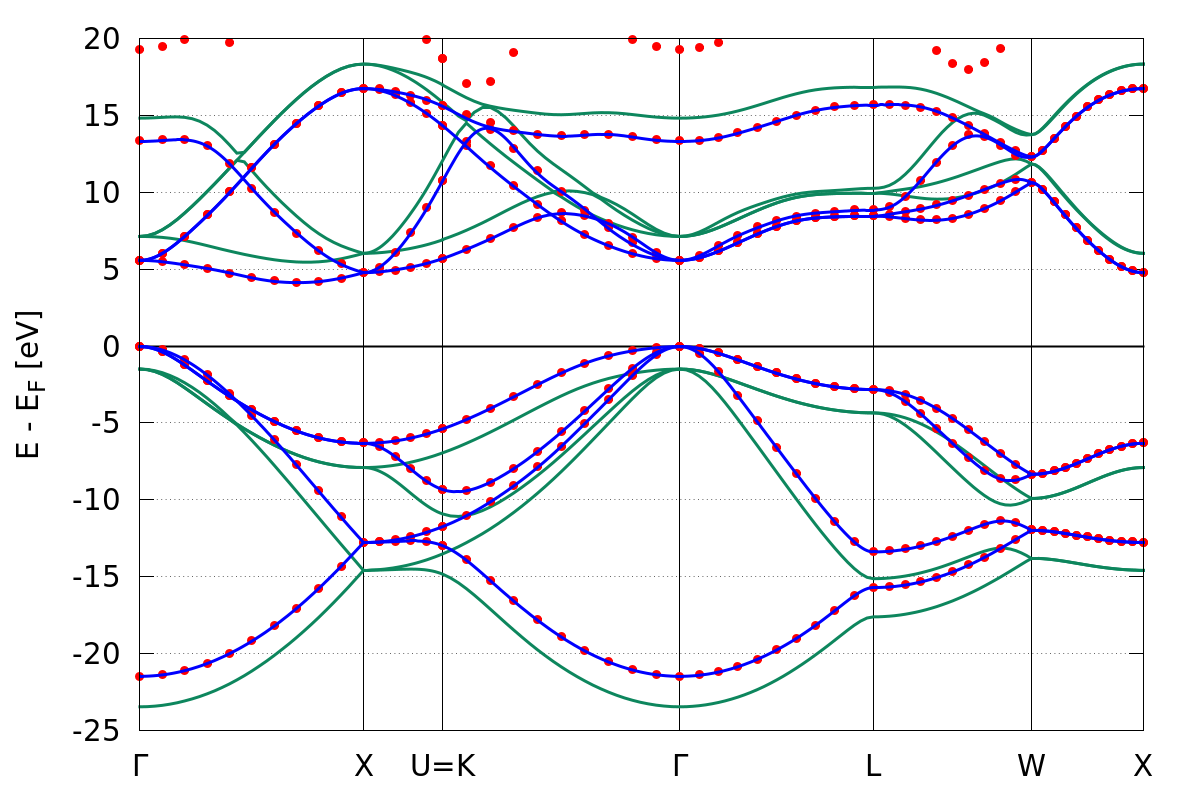}} \\
    \subfloat[BN: KI]{\includegraphics[width=.4\textwidth]{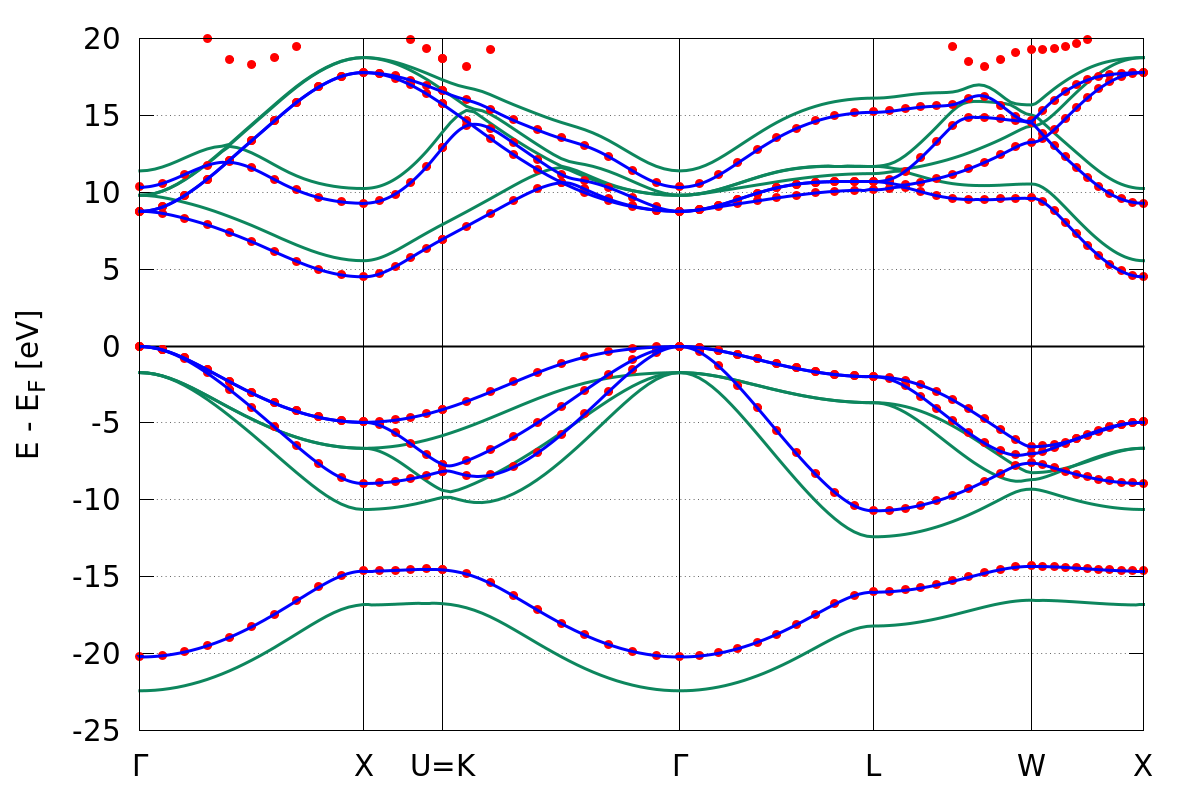}} \qquad
    \subfloat[BN: KIPZ]{\includegraphics[width=.4\textwidth]{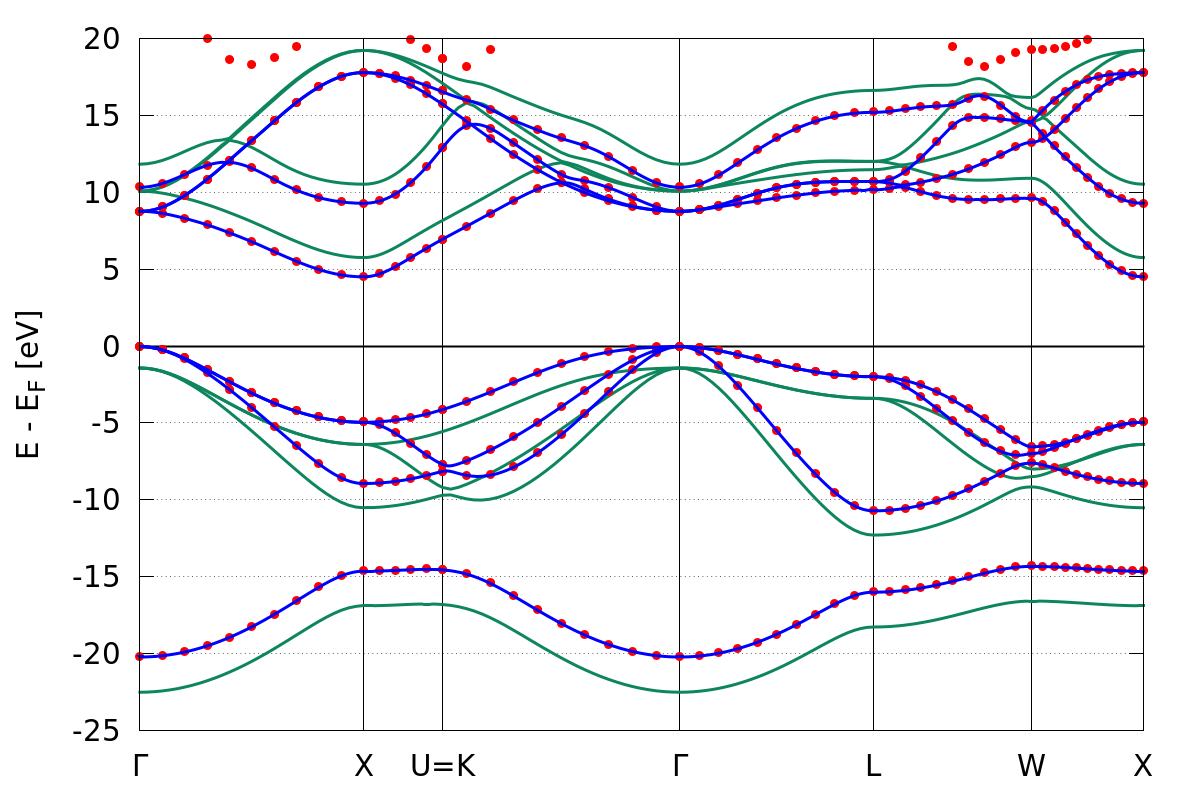}}
    \vspace{3mm}
    \begin{tabularx}{\linewidth}{*{4}{>{\centering\arraybackslash}X}}
    \hline
    \hline
    & Si & C & BN \\
    \hline
    occupied & $sp^3$ & $sp^3$ & N: $s$, $p$ \\
    empty    & $sp^3$ & $sp^3$ & B: $s$, $p$ \\
    \hline
    \end{tabularx}
    \caption{PBE (red circles) and Koopmans (green lines) band structure of Si, C and BN. The zero of the energy is set at the PBE Fermi level. The blue lines represent the PBE interpolated band structures as calculated by the Wannier90 code. In the table at the bottom we report the initial projections for the Wannierization procedure.}
    \label{fig:bands-1}
\end{figure*}

\begin{figure*}[htp]
    \subfloat[Ge: KI]{\includegraphics[width=.49\textwidth]{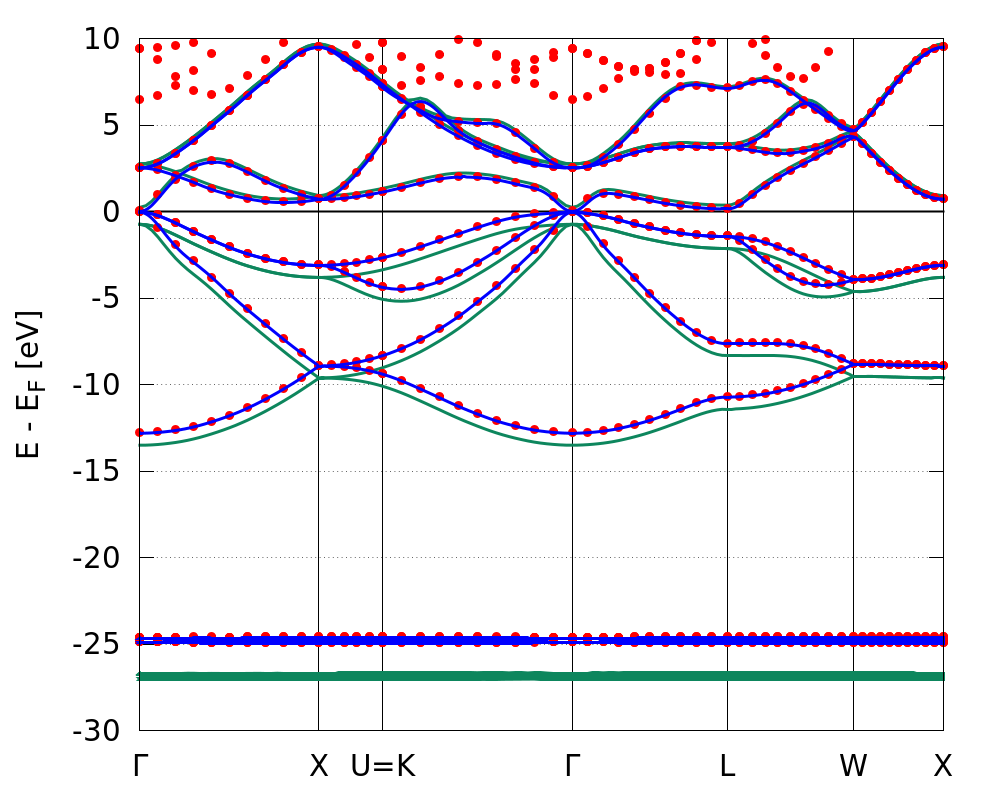}}
    \subfloat[Ge: KIPZ]{\includegraphics[width=.49\textwidth]{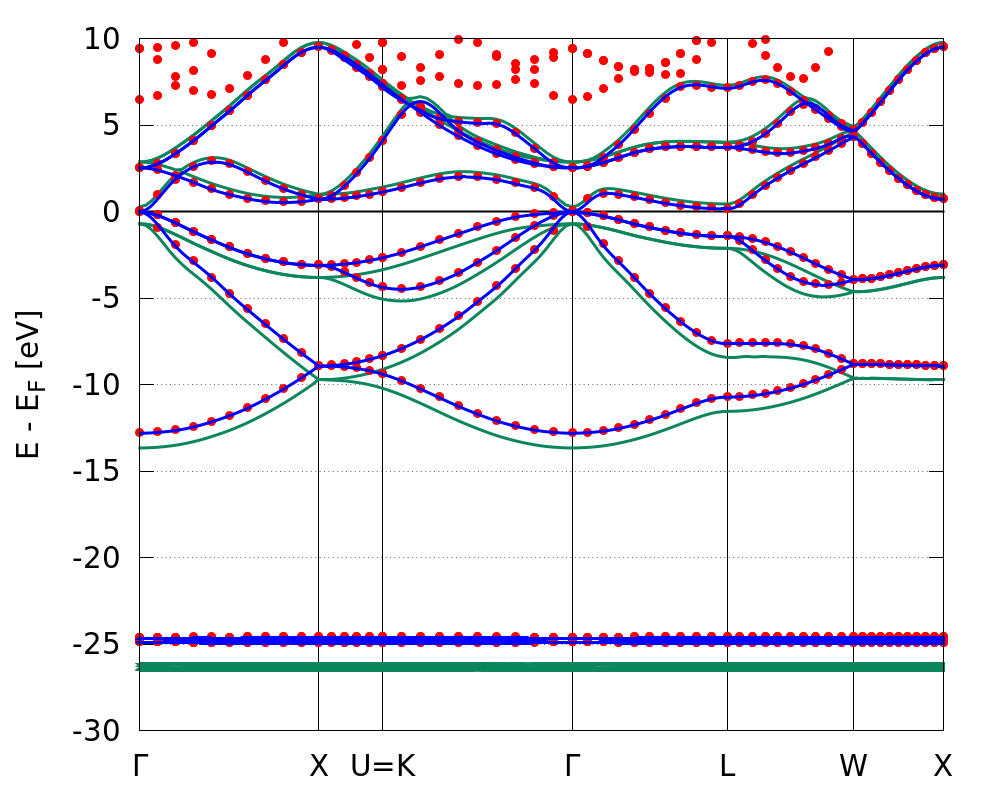}} \\
    \subfloat[GaAs: KI]{\includegraphics[width=.49\textwidth]{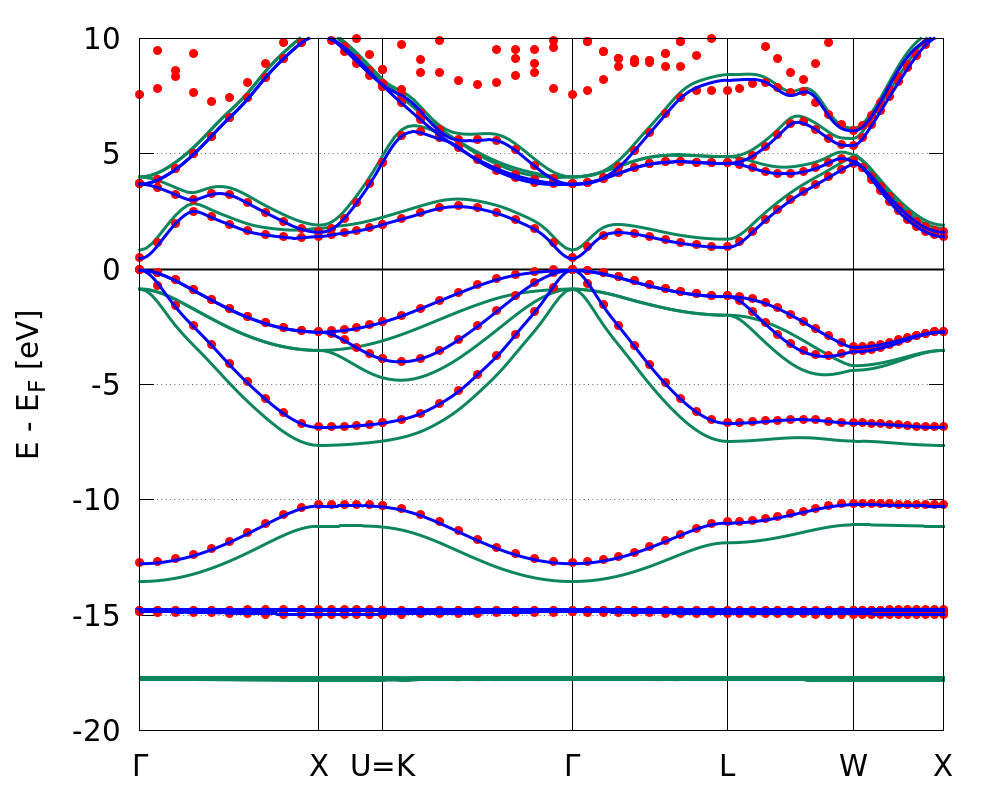}}
    \subfloat[GaAs: KIPZ]{\includegraphics[width=.49\textwidth]{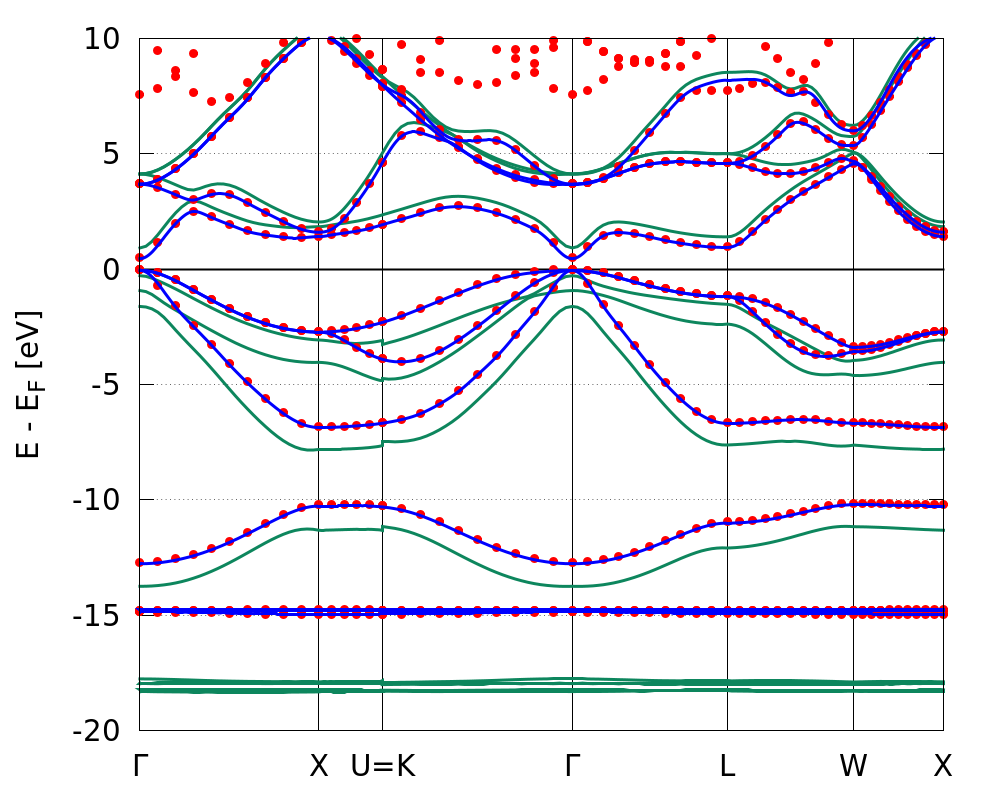}} \\
    \vspace{3mm}
    \begin{tabularx}{\linewidth}{*{3}{>{\centering\arraybackslash}X}}
    \hline
    \hline
    & Ge & GaAs \\
    \hline
    occupied & $d$, $d$, $sp^3$ & $d$, $sp^3$ \\
    empty    & $sp^3$           & $sp^3$      \\
    \hline
    \end{tabularx}
    \caption{As in Fig.~\ref{fig:bands-1}, here we report the PBE (blue lines) and KC (green lines) band structures of Ge and GaAs. The table at the bottom contains the initial projections for the Wannierization procedure.}
    \label{fig:bands-2}
\end{figure*}

\begin{figure*}[htp]
    \subfloat[MgO: KI]{\includegraphics[width=.47\textwidth]{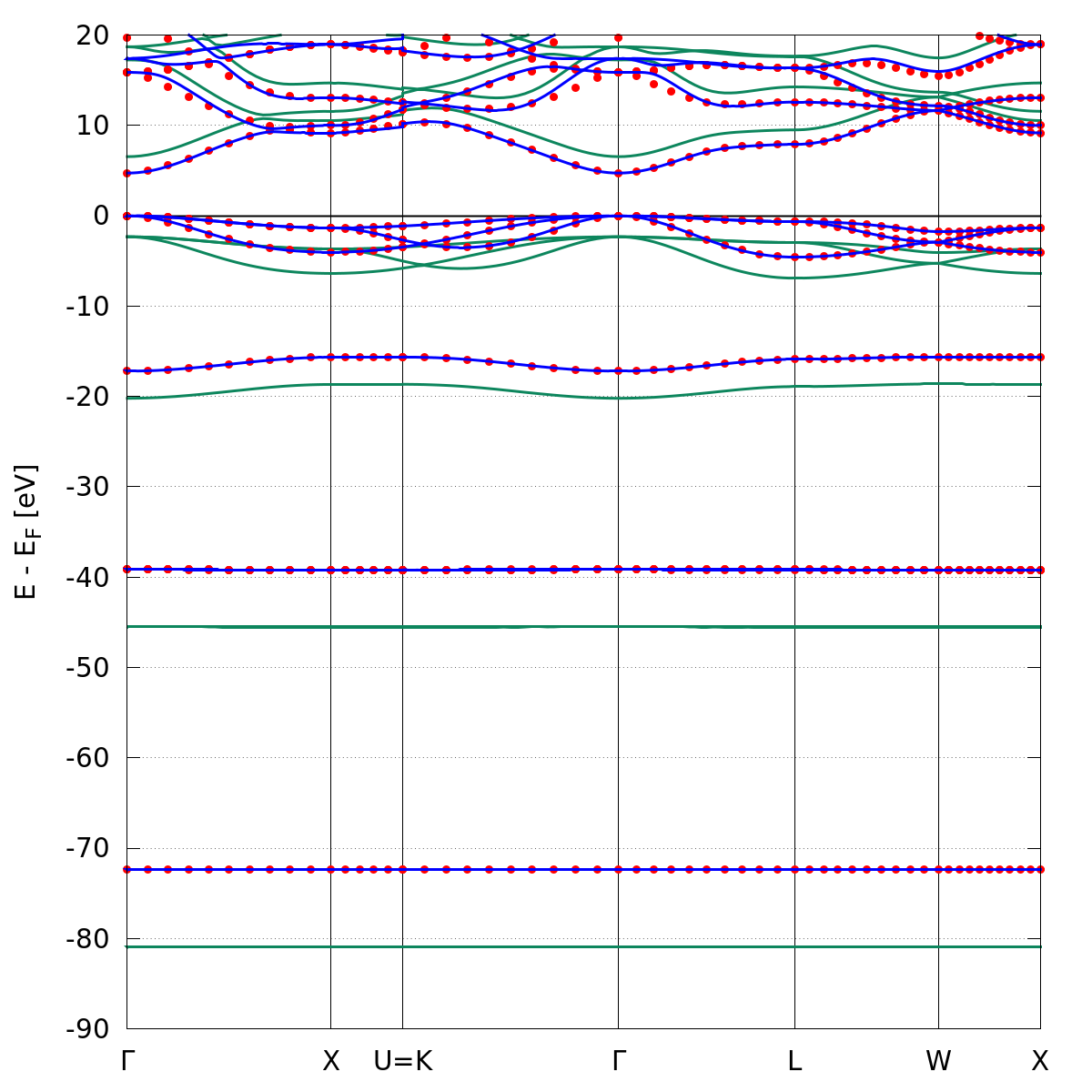}}
    \subfloat[MgO KIPZ]{\includegraphics[width=.47\textwidth]{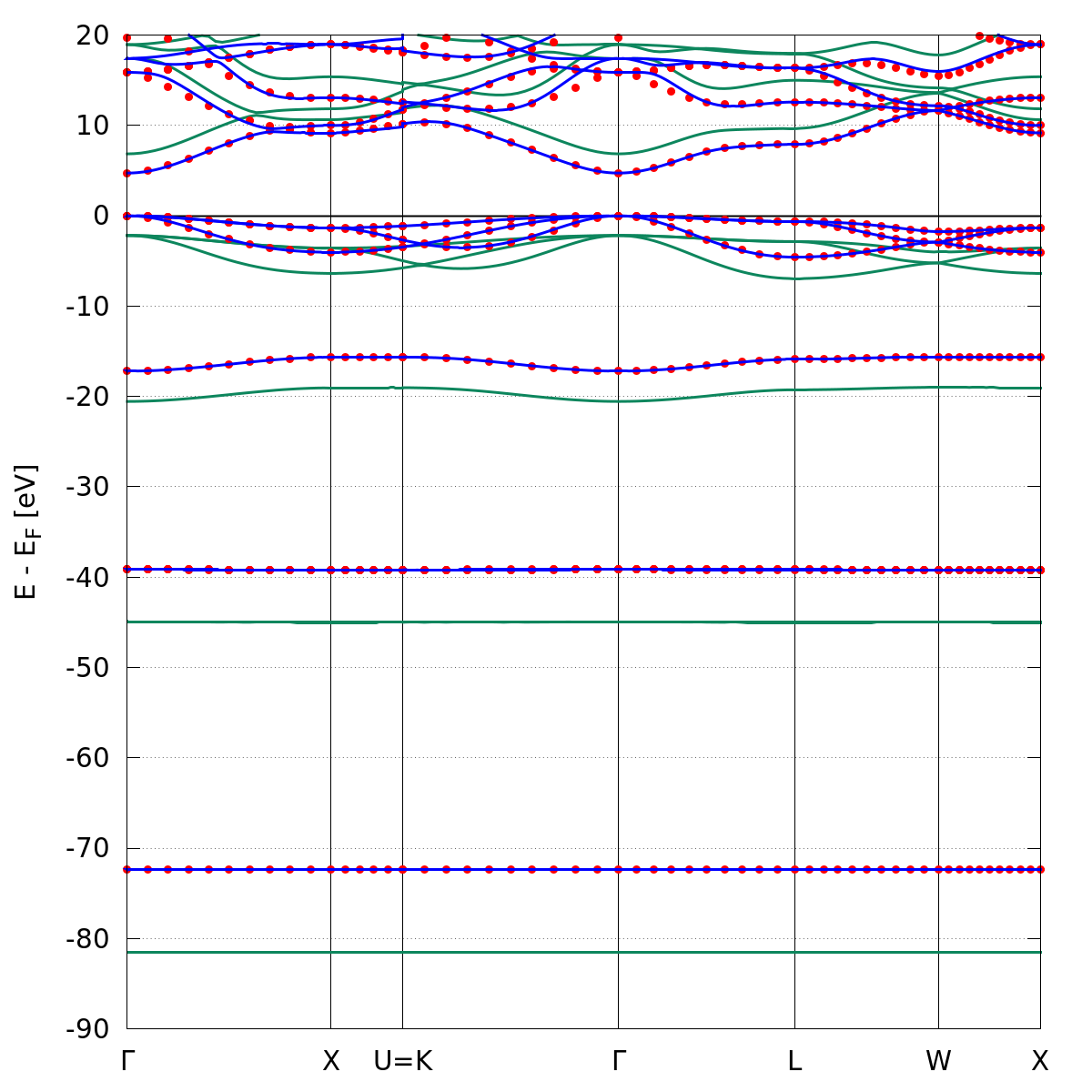}} \\
    \subfloat[LiF: KI]{\includegraphics[width=.47\textwidth]{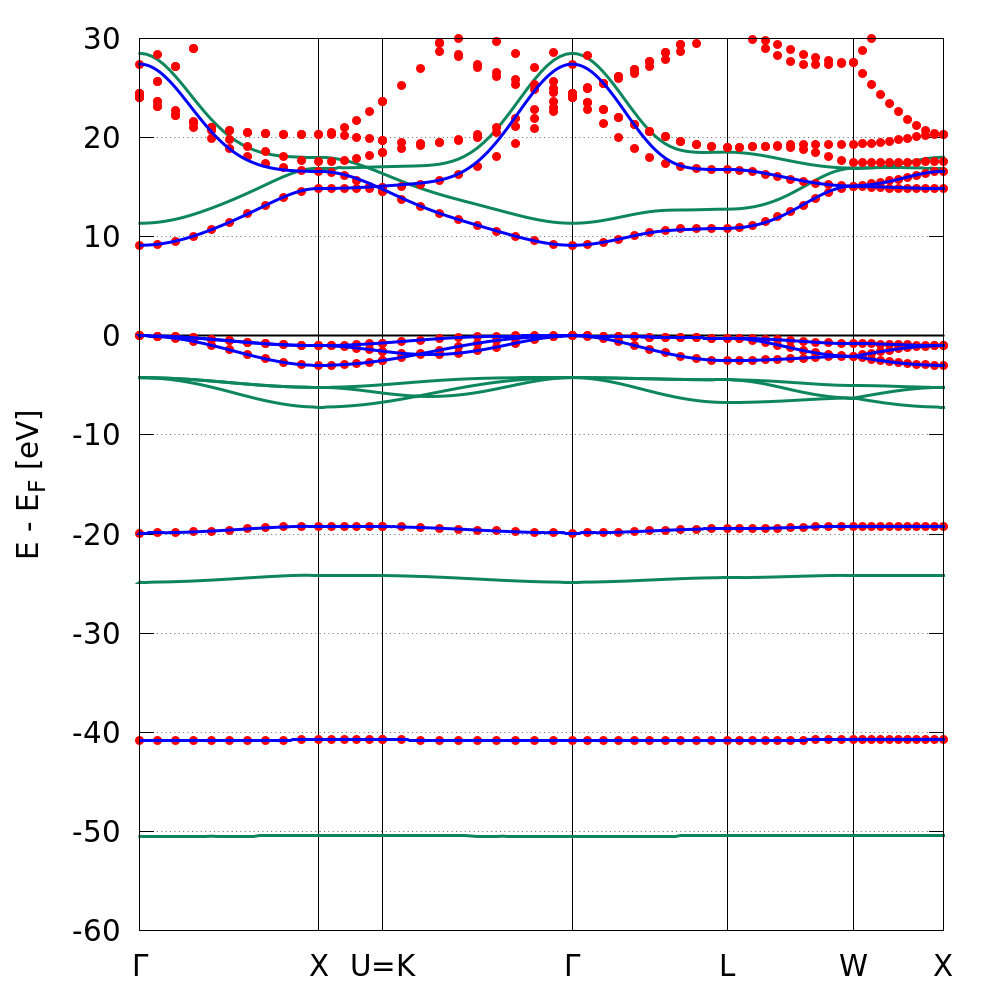}}
    \subfloat[LiF: KIPZ]{\includegraphics[width=.47\textwidth]{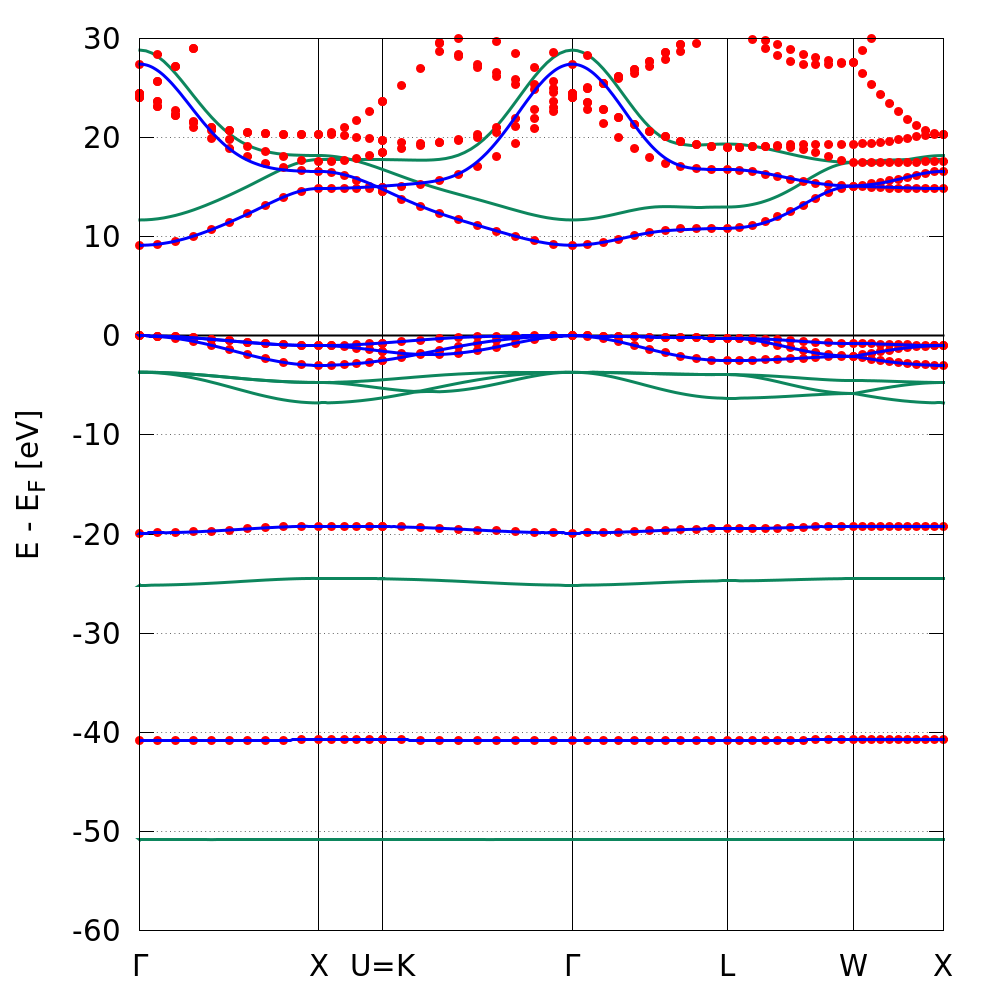}} \\
    \vspace{3mm}
    \begin{tabularx}{\linewidth}{*{3}{>{\centering\arraybackslash}X}}
    \hline
    \hline
    & MgO & LiF \\
    \hline
    occupied & Mg: $s$, $p$ / O: $s$, $p$ & Li: $s$, F: $s$, $p$ \\
    empty    & Mg: $s$ / O: $s$, $p$      & Li: $s$, F: $s$      \\
    \hline
    \end{tabularx}
    \caption{As in Fig.~\ref{fig:bands-1}, here we report the PBE (blue lines) and KC (green lines) band structures of MgO and LiF. The table at the bottom contains the initial projections for the Wannierization procedure.}
    \label{fig:bands-3}
\end{figure*}

We report in Figs.~\ref{fig:bands-1}-\ref{fig:bands-3} the band structures for bulk Ge, Si, GaAs, C-diamond, BN, MgO, and LiF. The Koopmans correction is very smooth with respect to $\bk$, thereby applying an almost constant shift (different for each group of bands) to the KS-DFT bands. As a consequence, the major contribution to the dispersion of the energy in $\bk$-space comes from the DFT part; in order to distinguish between possible variations due to a flawed interpolation or effectively caused by the KC correction, we report also the interpolated DFT band structure. This has been obtained using the Wannier90 code, that applies the same interpolation method explained in Sec.~\ref{sec:unfold-interpolate} to the DFT Hamiltonian only.

\begin{table*}[t]
    \centering
    \begin{tabularx}{\linewidth}{*{8}{>{\centering\arraybackslash}X}}
        \hline
        \hline
             &   PBE & QS$G\tilde{W}$ &    KI & pKIPZ &  KIPZ & Exp     &   ZPR \\
        \hline
        Si   &  0.55 &          1.24  &  1.18 &  1.17 &  1.19 &    1.17 &  $-0.05$ \\
        Ge   &  0.06 &          0.81  &  0.84 &  0.81 &  0.84 &    0.74 &  $-0.04$ \\
        GaAs &  0.50 &          1.61  &  1.53 &  1.49 &  1.50 &    1.52 &  $-0.05$ \\
        \hline
        C    &  4.16 &          5.90  &  6.48 &  6.51 &  6.58 &    5.50 &  $-0.36$ \\
        BN   &  4.52 &          6.59  &  6.83 &  6.67 &  6.73 &    6.20 &  $-0.42$ \\
        \hline
        MgO  &  4.73 &          8.30  &  8.23 &  8.04 &  8.40 &    7.83~\cite{shishkin_accurate_2007} &  $-0.64$ \\
        LiF  &  9.15 &         14.50  & 14.43 & 13.98 & 14.21 &   14.20~\cite{shishkin_accurate_2007} &  $-1.15$~\cite{nery_quasiparticles_2018} \\
        \hline
    \end{tabularx}
    \caption{Fundamental band gaps (in \si{\electronvolt}) as obtained from the three Koopmans flavours mentioned above; the reported band gaps include already the corrections due to zero-point effects. The values for Ge and GaAs are corrected by 0.1 $\si{\electronvolt}$ due to spin-orbit coupling. The QS$G\tilde{W}$ values are taken from Refs.~\cite{chen_accurate_2015,shishkin_accurate_2007}. If not specified otherwise, the experimental band gaps and the corresponding corrections due to zero-point motion effects are taken, respectively, from Refs. \cite{madelung_semiconductors_2004} and \cite{miglio_predominance_2020}. 
    }
    \label{tab:gaps}
\end{table*}

In Table \ref{tab:gaps} we report the band gaps obtained in this work. In most of the cases the agreement with the experiment is remarkable. In the case of diamond, all the Koopmans flavours significantly overestimate the gap of about 1~$\si{\electronvolt}$. The reason might be related to the high degree of localization of the variational orbitals (see Table~\ref{tab:sp3-sh}) that generally results in larger Koopmans corrections. We observe that, with respect to the other covalent semiconductors, the shift of the conduction bands of diamond is much larger. Using a different type of orbitals, e.g. starting from separate $s$- and $p$-like Wannier functions, reduces the band gap, thanks to the smaller localization of the orbitals; however, this requires further study and justification and it is not considered in this work. This case also unveils one of the limitations of the Koopmans approach: while the occupied variational orbitals are unambiguously defined as those that minimize the total energy, for the empty orbitals no analogous criterion exists. Here, the use of MLWFs represents a reasonable but nevertheless heuristic choice.

\begin{table}[b]
    \centering
    \begin{tabularx}{\linewidth}{*{3}{>{\centering\arraybackslash}X}}
        \hline
        \hline
             & occupied & empty \\
        \hline
        Si   &    6.219         & 2.980         \\
        Ge   &    5.757         & 3.120         \\
        GaAs &    6.068         & 3.093         \\
        C    &    9.344         & 5.925         \\
        BN   &    8.407 / 7.714 & 4.681 / 5.313 \\
        \hline
    \end{tabularx}
    \caption{Self-Hartree energies (in $\si{\electronvolt}$) of the (occupied and empty) MLWFs for the four covalent semiconductors and for boron nitride.}
    \label{tab:sp3-sh}
\end{table}

For Si, Ge, GaAs and BN the agreement with previously published Koopmans results \cite{nguyen_koopmans-compliant_2018} is within 0.1~$\si{\electronvolt}$. In the other cases the disagreement is probably due to the differences in the WFs used. We remark that, in Ref.~\cite{nguyen_koopmans-compliant_2018}, the whole procedure took place entirely within a supercell approach: the MLWFs were obtained from supercell $\Gamma$-sampling calculations where the additional degrees of freedom can give rise to qualitatively different Wannier functions. In fact, while in the supercell approach there is no constraint on the unitary transformation connecting Wannier and Bloch functions, in the primitive cell approach the unitary transformation is a block-diagonal (over $\bk$) matrix and so the mixing is different in the two approaches. This difference is even more marked when dealing with empty states: the presence of an entangled group of bands calls for a disentanglement procedure in order to define the Wannier functions, namely the selection of an optimal subset of Bloch states. This procedure is also $\bk$-dependent, thus the optimal set of Bloch functions selected in the supercell and primitive cell (even when the same energy windows are used) is generally different.

\begin{table}[t]
    \centering
    \begin{tabularx}{\linewidth}{*{1}{>{\arraybackslash}>{\hsize=2.0\hsize}X}*{6}{>{\centering\arraybackslash}X}}
        \hline
        \hline
                                                 &   PBE &  QS$GW$ &    KI &  KIPZ &   Exp  \\
        \hline
        Si & & & & & \\
        $\Gamma_{1v} \rightarrow \Gamma'_{25v}$  & 11.96 & 12.04 & 11.96 & 12.09 & 12.5(6)  \\
        $X_{4v} \rightarrow \Gamma'_{25v}$       &  2.84 &  2.99 &  2.84 &  2.86 &  2.9     \\
        $L_{1v} \rightarrow \Gamma'_{25v}$       &  6.96 &  7.18 &  6.96 &  7.04 &  6.8(2)  \\
        $L'_{2v} \rightarrow \Gamma'_{25v}$      &  9.63 &  9.79 &  9.63 &  9.74 &  9.3(4)  \\
        \\
        $\Gamma'_{25v} \rightarrow \Gamma_{15c}$ &  2.56 &  3.35 &  3.24 &  3.26 &  3.35(1) \\
        $\Gamma'_{25v} \rightarrow \Gamma'_{2c}$ &  3.33 &  4.08 &  4.00 &  4.01 &  4.15(1) \\
        $\Gamma'_{25v} \rightarrow X_{1c}$       &  0.69 &  1.44 &  1.36 &  1.37 &  1.13    \\
        $\Gamma'_{25v} \rightarrow L_{1c}$       &  1.51 &  2.27 &  2.18 &  2.19 &  2.04(6) \\
        $\Gamma'_{25v} \rightarrow L_{3c}$       &  3.33 &  4.24 &  3.99 &  4.00 &  3.9(1)  \\
        \hline
        GaAs & & & & & \\
        $\braket{\varepsilon_d}$                 & 14.9  & 17.6  & 16.9  & 17.7  & 18.9     \\
        \hline
        LiF & & & & & \\
        $\braket{\varepsilon_{1s({\rm Li})}}$    & 40.8  &  --   & 46.2  & 47.1  & 49.8     \\
        $\braket{\varepsilon_{2s({\rm F})}}$     & 19.5  &  --   & 20.2  & 21.0  & 23.9     \\
        \hline
    \end{tabularx}
    \caption{In the upper part we report the energy differences (in $\si{\electronvolt}$) for Si at special symmetry points in the BZ, wrt to the top of the valence band ($\Gamma'_{25v}$). The results from $GW$ calculations and the experimental values are taken, respectively, from Refs.~\cite{hybertsen_electron_1986} and \cite{madelung_semiconductors_2004}. The central part of the table contains the average position of the $d$-bands of GaAs; in this case the experimental and theoretical (self-consistent $GW$) values are taken from Ref.~\cite{shishkin_accurate_2007}. At the bottom we report the position of the $1s$ bands of Li and $2s$ bands of F in LiF; experimental values are taken from Ref.~\cite{johansson_core_1976}.}
    \label{tab:ene-special-points}
\end{table}

With regards to the rest of the spectrum, by looking at Figs.~\ref{fig:bands-1}-\ref{fig:bands-3} we see that the main effect of the Koopmans correction is a shift of the DFT band groups -- downward for the valence, upward for the conduction -- quite smoothly with respect to $\bk$, and fairly constant for bands corresponding to orbitals with the same chemical character. In principle, the only situation where the correction consists of a rigid shift of the bands is for the case of the KI potential acting on equivalent occupied states. From Eq.~\eqref{eq:ki-potential-occ}, we know that the KI potential for the occupied states is scalar, thus its representation on the variational orbitals yields a diagonal matrix. When the occupied orbitals have all the same character, which is for instance the case of the $sp^3$ orbitals in Si or C, the KI potential reduces to a multiple of the identity and the correction is a simple shift applied to all the DFT eigenvalues. On the other hand, in the presence of valence orbitals with a different character -- e.g. $sp^3$ and $d$ orbitals in GaAs -- although still diagonal, the KI potential is not anymore a multiple of the identity matrix and the effect of the correction is non-trivial. This becomes even more pronounced for the potential acting on the empty states or for the KIPZ potential: because of the presence of space-dependent terms, the matrix representation of the potential is non-diagonal and the Koopmans correction can affect also the bandwidth as well as the inter- and intra-band distances. Yet, the effect of these off-diagonal elements is minor: in a localized representation, such as that of the variational orbitals, and because of the local nature of the Koopmans Hamiltonian, the dominant matrix elements are diagonal, whereas the contribution from the off-diagonal matrix elements is second-order.

Finally, in Table~\ref{tab:ene-special-points} we can see that for Si the relative distance between valence bands (first block of points), well described already at the PBE level, is not modified by the KI correction. The second block of points shows the energies which are different from the fundamental band gap: the remarkable agreement with the results from photoemission experiments, and not only with the first ionization energies, further emphasizes the capability of KC functionals in predicting the full band structure.

\section{Conclusions}
Koopmans-compliant functionals have already proven to be an efficient and reliable orbital-density-dependent method for the calculation of spectral properties of materials, such as the ionization potentials and electron affinities of molecules, and the band gaps of solids. In this work we addressed the problems faced when calculating band structures of periodic materials with Koopmans-compliant functionals. When modeling electron addition/removal processes on top of local or semi-local functionals, the need for a localized set of orbitals forces us to rely on orbital densities that are not periodic over the primitive cell and appear to break the translation symmetry of the system. However, we showed that the use of Wannier functions as variational orbitals ensures the compliance of the Koopmans Hamiltonian -- as well as any other ODD Hamiltonian -- with Bloch's theorem, which allows us to describe the electronic states with a band-structure picture. We also showed how the dispersion in $\bk$-space of the electronic bands can be recovered from supercell $\Gamma$-sampling calculations thanks to an unfolding method that exploits the Wannier character of the orbitals and allows the interpolation of the band structure along any path within the BZ. This approach was then employed to calculate the band structures of prototypical bulk semi-conductors and insulators. These band structures showed remarkable agreement with experiment, displaying an accuracy comparable to more computationally intensive state-of-the-art many-body perturbation theory methods.

\begin{acknowledgments}
This work was supported by the Swiss National Science Foundation (SNSF) through the grant No. 200021-179138 and its National Centre of Competence in Research (NCCR) MARVEL.
\end{acknowledgments}

\appendix

\section{\label{app:kc-functionals-potentials}KI and KIPZ functionals and potentials}

The KI and KIPZ functionals are obtained, respectively, by applying the Koopmans ODD correction on top of a DFT functional (e.g. PBE) and of the PZ functional:
\begin{equation}
    E^{\rm KC}[\{ \rho_i \}] = E^{\rm PBE}[\rho] + \sum_i \alpha_i \Pi^{\rm uKC}_i[\rho,\rho_i]
    \label{eq:kc-functional-3}
\end{equation}
where in the case of the KI functional $\Pi^{\rm uKC}_i$ is given by Eq.~\eqref{eq:ki-correction-unrelaxed}, while for KIPZ we have
\begin{equation}
\Pi^{\rm uKIPZ}_i[\rho,\rho_i] = \Pi^{\rm uKI}_i[\rho,\rho_i] - f_i E_{\rm Hxc}[n_i] .
    \label{eq:kipz-functional}
\end{equation}
Note that in the one-particle limit there are no relaxation effects and the only $\alpha$ remaining is trivially equal to $1$; in this case, $E^{\rm KIPZ}[\{\rho_i\}]$ becomes exactly one-electron self-interaction free.

The variations of the functionals with respect to the orbital densities provide the action of the KC Hamiltonians:
\begin{equation}
    \frac{\delta E^{\rm KI}[\{ \rho_i \}]}{\delta \rho_j(\br)} = h^{\rm PBE}([\rho],\br) + \alpha_j v^{\rm KI}([\rho,\rho_j],\br)
    \label{eq:ki-Hamiltonian}
\end{equation}
and
\begin{equation}
    \frac{\delta E^{\rm KIPZ}[\{ \rho_i \}]}{\delta \rho_j(\br)} = h^{\rm PBE}([\rho],\br) + \alpha_j v^{\rm KIPZ}([\rho,\rho_j],\br) ,
    \label{eq:kipz-Hamiltonian}
\end{equation}
where $v^{\rm KI}([\rho_j],\br) = \sum_i \delta \Pi^{\rm uKI}_i / \delta \rho_j(\br)$ and the KIPZ potentials incorporate also the derivative of the additional self-interaction term $v^{\rm KIPZ}([\rho_j],\br) = v^{\rm KI}([\rho_j],\br) - \delta(f_j E_{\rm Hxc}[n_j]) / \delta \rho_j(\br)$. Finally, we report the expression of the KI potentials only for fully occupied ($f_j = 1$) and fully empty ($f_j = 0$) states, while for the most general expression given for any occupation number $f_j$ we refer to Ref.~\cite{borghi_koopmans-compliant_2014}:
\begin{equation}
\begin{split}
    v^{\rm KI,occ}[\rho,n_j] = \ & E_{\rm Hxc}[\rho] - E_{\rm Hxc}[\rho - n_j] \\
    & - \int d\br' v^{\rm Hxc}([\rho],\br') n_j(\br')
\end{split}
    \label{eq:ki-potential-occ}
\end{equation}
and
\begin{equation}
\begin{split}
    v^{\rm KI,emp}([\rho,n_j],\br) = \ & E_{\rm Hxc}[\rho + n_j] - E_{\rm Hxc}[\rho] \\
    & - \int d\br' v^{\rm Hxc}([\rho + n_j],\br') n_j(\br') \\
    & + \hat{v}^{\rm Hxc}([\rho + n_j],\br) - \hat{v}^{\rm Hxc}([\rho],\br) ,
\end{split}
    \label{eq:ki-potential-emp}
\end{equation}
\\
where we made use of $\delta \rho_i(\br') / \delta \rho_j(\br) = \delta_{ij} \delta(\br - \br')$ and of the fact that $\rho_j^{f_j=1}(\br) = n_j(\br)$.

By looking at Eq.~\eqref{eq:ki-potential-occ}, we see that the KI potential for the occupied states is fully scalar; the KI functional then, turns out to be unitary invariant and it shares the same ground-state density of the underlying DFT functional. On the other hand, the KIPZ potential always contains a non-scalar PZ term that is responsible for breaking of the unitary invariance (also with respect to transformations of the occupied subspace only) and, at the end of the minimization procedure, it allows to define without ambiguity the set of variational orbitals. In order to resolve the ambiguity affecting the KI functional, KI has been defined as the limit of the KIPZ functional when the PZ term goes to zero \cite{borghi_koopmans-compliant_2014}, which allows for a unique definition of the KI variational orbitals that are also determined by the $\alpha$PZ gradient (in this case the minimization consists only of an inner-loop since the KI ground-state density corresponds to the DFT one).

\section{\label{app:wannier-occupations}Occupation numbers of Wannier functions}

In this appendix we prove that the occupation numbers of Wannier functions are independent of the lattice vector $\bR$.

In terms of the KS (Bloch-like) eigenstates, the total electronic density is $\rho(\br) = \sum_{\bk,n} f_{\bk n} \psi^*_{\bk n}(\br) \psi_{\bk n}(\br)$ where the occupations $f_{\bk n}$ follow Fermi-Dirac statistics. We now consider the transformation connecting Bloch and Wannier functions \cite{marzari_maximally_2012}:
\begin{equation}
    w_{\bR m}(\br) = \frac{V}{(2\pi)^3} \int_{BZ} d\bk e^{-i\bk \cdot \bR} \sum_m U^{(\bk)}_{nm} \psi_{\bk n}(\br)
    \label{eq:wannier-bloch}
\end{equation}
which inverted gives
\begin{equation}
    \psi_{\bk n}(\br) = \sum_{\bR, m} e^{i\bk \cdot \bR} U^{(\bk)*}_{nm} w_{\bR m}(\br) .
    \label{eq:bloch-wannier}
\end{equation}
On the Wannier basis the density takes the form
\begin{equation}
    \rho(\br) = \sum_{\bR,\bR',m,n} f_{mn}^{\bR\bR'} w^*_{\bR m}(\br) w_{\bR' n}(\br) , 
\end{equation}
where $f_{mn}^{\bR\bR'} = \sum_{\bk p} f_{\bk p} e^{-i\bk(\bR-\bR')} U^{(\bk)}_{pm} U^{(\bk)*}_{pn}$. Therefore, the matrix elements $f_{mn}^{\bR\bR'}$ depend only on the difference between $\bR$ and $\bR'$:
\begin{equation}
    f_{mn}^{\bR\bR'} = f_{mn}^{\bR-\bR'} .
    \label{eq:matrix-elements-occupations-wannier}
\end{equation}

As a consequence of Eq.~\eqref{eq:matrix-elements-occupations-wannier} the occupancies on the Wannier orbitals, i.e. the diagonal elements of the matrix $f_{mn}^{\bR\bR'}$, are independent from the lattice vector:
\begin{equation}
    f_{\bR n} = f_{nn}^{\bR\bR} = f_{nn}^{\bR-\bR} = f_{nn}^{\bm 0} = f_{{\bm 0}n},
\end{equation}
as claimed.

\section{\label{app:bloch-theorem}Commutativity of \texorpdfstring{$\hat{v}^{\rm PZ}$}{} with the translation operators}

In this appendix we demonstrate that the PZ potential $\hat{v}^{\rm PZ}$, as defined in Eq.~\eqref{eq:pz-potential}, commutes with a primitive-cell translation operator $\hat T_\bR$. To do so, let us consider the action of the PZ potential on a generic state $\psi$ and start by projecting onto position space:
\begin{equation}
    \begin{split}
    v^{\rm PZ}(\br) \psi(\br) &= \braket{\br | \hat{v}^{\rm PZ} | \psi} \\
    &= \bra{\br} \sum_{\bR'} - \hat{v}^{\rm Hxc}[\rho_{\bR'}] \ket{w_{\bR'}} \braket{w_{\bR'} | \psi} \\
    &= \sum_{\bR'} - v^{\rm Hxc}([\rho_{\bR'}],\br) w_{\bR'}(\br) \braket{w_{\bR'} | \psi} ;
    \end{split}
    \label{eq:projection-real-space}
\end{equation}
we then apply a given translation operator (over the primitive cell lattice vectors $\bR$) $\hat{T}_{\bR}: f(\br) \longmapsto f(\br + \bR)$ to Eq.~\eqref{eq:projection-real-space}:
\begin{equation}
    \begin{split}
    \hat{T}_{\bR} v^{\rm PZ}(\br) \psi(\br) &= - \sum_{\bR'} \hat{T}_{\bR} v^{\rm Hxc}([\rho_{\bR'}],\br) w_{\bR'}(\br) \braket{w_{\bR'} | \psi} \\
    &= - \sum_{\bR'} v^{\rm Hxc}([\rho_{\bR'}],\br+\bR) w_{\bR'}(\br+\bR) \braket{w_{\bR'} | \psi} \\
    &= - \sum_{\bR'} v^{\rm Hxc}([\rho_{\bR'-\bR}],\br) w_{\bR'-\bR}(\br) \braket{w_{\bR'} | \psi}
    \end{split}
    \label{eq:bloch-th-1}
\end{equation}
where we used Eqs.~\eqref{eq:trans-prop-wannier} and \eqref{eq:trans-prop-pot}. By changing variables $\bR'-\bR \longrightarrow \bR'$ we obtain
\begin{equation}
    \hat{T}_{\bR} v^{\rm PZ}(\br) \psi(\br) = - \sum_{\bR'} v^{\rm Hxc}([\rho_{\bR'}],\br) w_{\bR'}(\br) \braket{w_{\bR'+\bR} | \psi} .
    \label{eq:bloch-th-2}
\end{equation}
The overlap $\braket{w_{\bR'+\bR} | \psi}$ can then be rewritten as follows:
\begin{equation}
    \begin{split}
    \braket{w_{\bR'+\bR} | \psi} &= \int d\br'\ w_{\bR'+\bR}^*(\br') \psi(\br') \\
    &= \int d\br'\ w_{\bR'}^*(\br') \psi(\br'+\bR) \\
    &= \bra{w_{\bR'}} \int d\br'\ \ket{\br'} \psi(\br'+\bR) .
    \end{split}
    \label{eq:overlap-psi-w}
\end{equation}

By putting together Eqs.~\eqref{eq:bloch-th-2} and \eqref{eq:overlap-psi-w} and expressing $v^{\rm Hxc}([\rho_{\bR'}],\br) w_{\bR'}(\br)$ as a projection on position space, we obtain
\begin{equation}
    \begin{split}
    \hat{T}_{\bR} v^{\rm PZ}(\br) \psi(\br) &= - \bra{\br} \sum_{\bR'} \hat{v}^{\rm Hxc}[\rho_{\bR'}] \ket{w_{\bR'}} \bra{w_{\bR'}} \int d\br'\ \ket{\br'} \psi(\br'+\bR) \\
    &= \int d\br'\ \braket{\br|\hat{v}^{\rm PZ}|\br'} \psi(\br'+\bR) \\
    &= v^{\rm PZ}(\br) \psi(\br+\bR) \\
    &= v^{\rm PZ}(\br) \hat{T}_{\bR} \psi(\br) ,
    \end{split}
    \label{eq:bloch-th-3}
\end{equation}
where we used the locality of the PZ potential, $\braket{\br|\hat{v}^{\rm PZ}|\br'} = v^{\rm PZ}(\br) \delta(\br-\br')$.

\section{\label{app:smooth-interpolation}Smooth interpolation method}

When reconstructing the band structure from a supercell calculation, one faces a trade-off: on one hand, a sufficiently large supercell must be used to minimize the errors associated with neglecting long-range matrix elements of the Hamiltonian (see Sec.~\ref{sec:unfold-interpolate}). But on the other hand, increasing the size of the supercell dramatically increases the computational costs. In this scenario, one can exploit the fact that the KC potential is a small, slowly varying correction on top of the original DFT Hamiltonian. If one decomposes the right-hand side of Eq.~\eqref{eq:hk-unfold} in its DFT and KC components,
\begin{equation}
    h^{\rm KC}_{mn}(\bR) \longrightarrow h^{\rm DFT}_{mn}(\bR) + v^{\rm KC}_{mn}(\bR) ,
\end{equation}
it is reasonable to assume that the major source of error comes from the interpolation of $h^{\rm DFT}_{mn}(\bR)$. This allows to improve the interpolation of the band structure by rewriting Eq.~\eqref{eq:hk-unfold} as
\begin{equation}
    h^{\rm KC}_{mn}(\bk) = \sum_{\bR'} e^{i \bk \cdot \bR'} h^{\rm DFT}_{mn}(\bR') + \sum_{\bR} e^{i \bk \cdot \bR} v^{\rm KC}_{mn}(\bR)
    \label{eq:hk-smooth}
\end{equation}
where the set of vectors $\{ \bR' \}$ now corresponds to a much larger supercell or, equivalently, it comes from a calculation with a denser $\bk$-points grid. This represents a significant saving in computational costs because the Koopmans calculations can be then performed on smaller supercells than would otherwise be necessary.

In order to have a consistent representation between (a) the DFT Hamiltonian defined  on a very dense grid, and (b) the KC potential on a coarser grid, it is important to have the same set of WFs for the two calculations. As long as this is fulfilled, the Koopmans Hamiltonian can be factorized as shown in Eq.~\eqref{eq:hk-smooth} and the DFT part can be obtained starting from a $\bk$-points grid dense enough to reliably interpolate the band structure.

We stress that this method has the one goal of improving the interpolation of the band structure. The convergence of other results, such as total energies and eigenvalues on the $\bk$-points grid commensurate with the supercell, is typically achieved with relatively small supercells. The technique depicted above does not affect any of these quantities and only improves the results for the electronic eigenvalues at $\bk$-points not included in the original Monkhorst-Pack grid.

\bibliographystyle{apsrev4-2}
\bibliography{main}

\end{document}